\documentclass[11pt,twoside]{article}
\usepackage{amsmath}
\usepackage{amssymb}
\usepackage{amscd}
\newcommand{\remark}{\smallbreak\noindent{\bf Remark.}}
\renewcommand{\a}{\alpha}
\renewcommand{\b}{\beta}
\newcommand{\g}{\gamma}
\newcommand{\G}{\Gamma}
\renewcommand{\d}{\delta}
\newcommand{\e}{\varepsilon}
\renewcommand{\k}{\kappa}
\renewcommand{\l}{\lambda}
\newcommand{\La}{\Lambda}
\newcommand{\m}{\mu}

\newcommand{\s}{\sigma}
\renewcommand{\t}{\tau}
\newcommand{\vph}{\varphi}
\newcommand{\om}{{\omega}}
\newcommand{\Cs}{
   {\rlap{\lower3pt\hbox{\textnormal{\LARGE\char'040}}}{\Gamma}}{}}
\newcommand{\de}{\partial}
\newcommand{\Eo}{{\scriptstyle{\mathrm{E}}}}
\newcommand{\oh}{\tfrac{1}{2}}
\newcommand{\osq}{\tfrac{1}{\surd2}}
\newcommand{\cj}[1]{\overline{#1}}
\newcommand{\lin}{{\scriptscriptstyle\bigstar}}
\newcommand{\alin}{{\overline{\scriptscriptstyle\bigstar}}}
\renewcommand{\.}{{\scriptstyle\boldsymbol{\dot{}}}}
\newcommand{\td}{\tilde}
\newcommand{\br}{\breve}
\newcommand{\gf}{g^\diamond}
\newcommand{\interaction}{{}_{\sst{\mathrm{int}}}}
\newcommand{\sbot}{{\scriptscriptstyle\bot}}
\newcommand{\sbo}{{\!\sbot}}
\newcommand{\fl}{\flat}
\newcommand{\bvv}{{\bar\vv}}
\newcommand{\B}{{\boldsymbol{B}}}

\newcommand{\Bb}{{\scriptscriptstyle{\boldsymbol{B}}}}
\newcommand{\E}{{\boldsymbol{E}}}
\newcommand{\F}{{\boldsymbol{F}}}
\newcommand{\Ft}{{\tilde\F}}
\renewcommand{\H}{{\boldsymbol{H}}}
\newcommand{\K}{{\boldsymbol{K}}}
\newcommand{\M}{{\boldsymbol{M}}}
\newcommand{\Mm}{{\scriptscriptstyle{\boldsymbol{M}}}}
\renewcommand{\P}{{\boldsymbol{P}}}
\newcommand{\Pm}{\P_{\!\!m}}
\newcommand{\Px}[1]{\P_{\!\!#1}}
\newcommand{\Pz}{\P_{{\!\!}_0}}
\newcommand{\Ptru}{\P\!\!_{\sst\vartriangle}}
\newcommand{\T}{{\boldsymbol{T}}}
\newcommand{\Tt}{{\scriptscriptstyle{\boldsymbol{T}}}}
\newcommand{\V}{{\boldsymbol{V}}}
\newcommand{\Vc}{\cj{\V}}
\newcommand{\Vl}{\V{}^\lin}
\newcommand{\W}{{\boldsymbol{W}}}
\newcommand{\Wc}{\cj{\W}}
\newcommand{\Wl}{\W{}^\lin}
\newcommand{\X}{{\boldsymbol{X}}}
\newcommand{\Xx}{{\scriptscriptstyle{\boldsymbol{X}}}}
\newcommand{\Y}{{\boldsymbol{Y}}}
\newcommand{\Z}{{\boldsymbol{Z}}}
\newcommand{\CC}{{\mathbb{C}}}
\newcommand{\LL}{{\mathbb{L}}}
\newcommand{\NN}{{\mathbb{N}}}
\newcommand{\RR}{{\mathbb{R}}}
\newcommand{\VV}{{\mathbb{V}}}
\newcommand{\CCc}{{\scriptscriptstyle{\mathbb{C}}}}
\newcommand{\RRr}{{\scriptscriptstyle{\mathbb{R}}}}
\newcommand{\Acal}{{\mathcal{A}}}
\newcommand{\Bcal}{{\mathcal{B}}}
\newcommand{\Dcal}{{\mathcal{D}}}
\newcommand{\Fcal}{{\mathcal{F}}}
\newcommand{\Hcal}{{\mathcal{H}}}
\newcommand{\Ical}{{\mathcal{I}}}
\newcommand{\Pcal}{{\mathcal{P}}}
\newcommand{\Scal}{{\mathcal{S}}}
\newcommand{\Ucal}{{\mathcal{U}}}
\newcommand{\Vcal}{{\mathcal{V}}}
\newcommand{\Ycal}{{\mathcal{Y}}}
\newcommand{\Cfr}{\mathfrak{C}}
\newcommand{\Gfr}{\mathfrak{G}}
\newcommand{\Hfr}{\mathfrak{H}}
\newcommand{\AC}{{\boldsymbol{\Acal}}}
\newcommand{\BC}{{\boldsymbol{\Bcal}}}
\newcommand{\DC}{{\boldsymbol{\Dcal}}}
\newcommand{\FC}{{\boldsymbol{\Fcal}}}
\newcommand{\HC}{{\boldsymbol{\Hcal}}}
\newcommand{\IC}{{\boldsymbol{\Ical}}}
\newcommand{\PC}{{\boldsymbol{\Pcal}}}
\newcommand{\VC}{{\boldsymbol{\Vcal}}}

\newcommand{\YC}{{\boldsymbol{\Ycal}}}
\newcommand{\BCo}{\BC_{\!\circ}}
\newcommand{\DCo}{\DC_{\!\circ}}
\newcommand{\DCh}{{\rlap{\;/}\DC}}

\newcommand{\FCo}{\FC_{\!\circ}}
\newcommand{\VCo}{{\VC\!_{\circ}}}
\newcommand{\YCo}{{\YC_{\!\circ}}}
\newcommand{\End}{\operatorname{End}}
\newcommand{\Ker}{\operatorname{Ker}}
\newcommand{\Id}[1]{{1\!\!1}\!{}_{#1}{}}
\newcommand{\id}{{1\!\!1}}
\newcommand{\dO}{\mathrm{d}}
\newcommand{\LO}{\mathrm{Lin}}
\newcommand{\he}{{\scriptstyle{\mathrm{H}}}}
\newcommand{\JO}{\mathrm{J}}
\newcommand{\kO}{\mathrm{k}}
\newcommand{\TO}{\mathrm{T}}
\newcommand{\TS}{\TO^{*}\!}
\newcommand{\VO}{\mathrm{V}}
\newcommand{\VS}{\VO^{*}\!}
\newcommand{\dt}{\dO\tt}
\newcommand{\eO}{\mathrm{e}}
\newcommand{\iO}{\mathrm{i}}

\newcommand{\ten}[1]{\operatorname*{\otimes}_{\!{\scriptscriptstyle #1}} }
\newcommand{\cart}[1]{\operatorname*{\times}_{\!{\scriptscriptstyle #1}} }
\newcommand{\dir}[1]{\operatorname*{\oplus}_{\!{\scriptscriptstyle #1}} }
\newcommand{\we}{{\,\wedge\,}}
\newcommand{\weu}[1]{{\wedge^{\!#1}}}
\newcommand{\comp}{\mathbin{\raisebox{1pt}{$\scriptstyle\circ$}}}
\newcommand{\tn}{{\,\otimes\,}}
\newcommand{\ve}{{\,\vee\,}}
\newcommand{\bang}[1]{{\langle#1\rangle}}
\newcommand{\topf}[1]{{\upharpoonleft}#1{\upharpoonright}}
\newcommand{\Ii}[2]{{}^{#1}_{\phantom{#1}\!#2}}
\newcommand{\iIi}[3]{{}_{#1\phantom{#2}\!\!#3}^{\phantom{#1}\!#2}}
\newcommand{\IiI}[3]{{}^{#1\phantom{#2}\!\!#3}_{\phantom{#1}\!#2}}
\newcommand{\sA}{{\scriptscriptstyle A}}
\newcommand{\sB}{{\scriptscriptstyle B}}
\newcommand{\sQ}{{\scriptscriptstyle Q}}
\newcommand{\cA}{{\sA\.}}
\newcommand{\cB}{{\sB\.}}
\newcommand{\Asf}{{\mathsf{A}}}
\newcommand{\Bsf}{{\mathsf{B}}}
\newcommand{\bb}{{\mathsf{b}}}
\newcommand{\Csf}{{\mathsf{C}}}
\newcommand{\Ysf}{{\mathsf{Y}}}
\newcommand{\yYsf}{{\sst\Ysf}}
\newcommand{\ee}{{\mathsf{e}}}
\newcommand{\kk}{{\mathsf{k}}} \newcommand{\ks}{\kk_\sbo}
\newcommand{\pp}{{\mathsf{p}}} \newcommand{\ps}{\pp_\sbo}
\newcommand{\qq}{{\mathsf{q}}} \newcommand{\qs}{\qq_\sbo}
\renewcommand{\tt}{{\mathsf{t}}}
\newcommand{\uu}{{\mathsf{u}}}
\newcommand{\vv}{{\mathsf{v}}}
\newcommand{\xx}{{\mathsf{x}}}
\newcommand{\yy}{{\mathsf{y}}}
\newcommand{\ie}{i.e$.$}
\newcommand{\sst}{\scriptscriptstyle}
\newcommand{\ul}{\underline}
\newcommand{\tps}{time\raise1pt\hbox{$\scriptscriptstyle{{+}}$}space}
\newcommand{\VRwa}{\qbezier(0,0)(1,1)(0,2)}

\newcommand{\VLwa}{\qbezier(0,0)(-1,1)(0,2)}
\newcommand{\Vwa}{\qbezier(0,0)(0,1)(0,2)}
\newcommand{\EVRwa}[1]{\multiput(0,0)(0,4){#1}{\VRwa \put(0,2){\VLwa} }}
\newcommand{\OVRwa}[1]{\VRwa
   \multiput(0,2)(0,4){#1}{\VLwa \put(0,2){\VRwa} }}
\newcommand{\EVLwa}[1]{\multiput(0,0)(0,4){#1}{\VLwa \put(0,2){\VRwa} }}
\newcommand{\OVLwa}[1]{\VLwa
   \multiput(0,2)(0,4){#1}{\VRwa \put(0,2){\VLwa} }}
\newcommand{\HRwa}{\qbezier(0,0)(1,1)(2,0)}
\newcommand{\HLwa}{\qbezier(0,0)(1,-1)(2,0)}
\newcommand{\EHRwa}[1]{\multiput(0,0)(4,0){#1}{\HRwa \put(2,0){\HLwa} }}
\newcommand{\DeLwa}{\qbezier(0,0)(.1,1.4)(1.5,1.5)}
\newcommand{\DeRwa}{\qbezier(0,0)(1.4,.1)(1.5,1.5)}
\newcommand{\EDeLwa}[1]{
   {\multiput(0,0)(3,3){#1}{\DeLwa \put(1.5,1.5){\DeRwa} }}}
\newcommand{\ODeRwa}[1]{\DeRwa
   {\multiput(1.5,1.5)(3,3){#1}{\DeLwa \put(1.5,1.5){\DeRwa} }}}
\newcommand{\DwLwa}{\qbezier(0,0)(-1.4,.1)(-1.5,1.5)}
\newcommand{\DwRwa}{\qbezier(0,0)(-.1,1.4)(-1.5,1.5)}
\newcommand{\EDwLwa}[1]{
   {\multiput(0,0)(-3,3){#1}{\DwLwa \put(-1.5,1.5){\DwRwa} }}}
\newcommand{\EDwRwa}[1]{
   {\multiput(0,0)(-3,3){#1}{\DwRwa \put(-1.5,1.5){\DwLwa} }}}
\newcommand{\ODwRwa}[1]{\DwRwa
   {\multiput(-1.5,1.5)(-3,3){#1}{\DwLwa \put(-1.5,1.5){\DwRwa} }}}
\newcommand{\ZeRwa}{\qbezier(0,0)(0.8825,0.81)(0.485,1.94)}
\newcommand{\ZeLwa}{\qbezier(0,0)(-0.3975,1.13)(0.485,1.94)}
\newcommand{\EZeRwa}[1]{
   {\multiput(0,0)(.97,3.88){#1}{\ZeRwa \put(.485,1.94){\ZeLwa} }}}
\newcommand{\ZwRwa}{\qbezier(0,0)(-0.8825,0.81)(-0.485,1.94)}
\newcommand{\ZwLwa}{\qbezier(0,0)(0.3975,1.13)(-0.485,1.94)}
\newcommand{\EZwRwa}[1]{
   {\multiput(0,0)(-.97,3.88){#1}{\ZwRwa \put(-.485,1.94){\ZwLwa} }}}
\newcommand{\Lcqs}{
\begin{picture}(20,20)(0,5) \linethickness{.3pt}
\put(10,20){\vector(-1,-2){7}} \put(5,10){\line(-1,-2){5}}
\put(20,0){\vector(-1,2){7}} \put(15,10){\line(-1,2){5}}
\put(9.8,0){\OVLwa{4} \put(0,18){\Vwa} }
\end{picture} }
\newcommand{\LSqs}{
\begin{picture}(20,20)(0,5) \linethickness{.3pt}
\put(10,10){\vector(-1,1){7}} \put(10,10){\line(-1,1){10}}
\put(20,0){\vector(-1,1){7}} \put(20,0){\line(-1,1){10}}
\put(10,-.2){\OVLwa{2}}
\end{picture} }
\newcommand{\LcQs}{
\begin{picture}(20,20)(0,5) \linethickness{.3pt}
\put(10,10){\vector(-1,-1){7}} \put(10,10){\line(-1,-1){10}}
\put(20,0){\vector(-1,1){7}} \put(20,0){\line(-1,1){10}}
\put(10,10){\OVLwa{2}}
\end{picture} }
\newcommand{\LSQs}{
\begin{picture}(20,20)(0,5) \linethickness{.3pt}
\put(20,0){\vector(-1,1){7}} \put(20,0){\line(-1,1){10}}
\put(10,10){\vector(-1,1){7}} \put(10,10){\line(-1,1){10}}
\put(10,10){\OVLwa{2}}
\end{picture} }
\newcommand{\LSqC}{
\begin{picture}(20,20)(0,5) \linethickness{.3pt}
\put(10,10){\vector(-1,1){7}} \put(10,10){\line(-1,1){10}}
\put(20,20){\vector(-1,-1){7}} \put(20,20){\line(-1,-1){10}}
\put(10,-.2){\OVLwa{2}}
\end{picture} }
\newcommand{\LSQC}{
\begin{picture}(20,20)(0,5) \linethickness{.3pt}
\put(10,0){\vector(-1,2){7}} \put(5,10){\line(-1,2){5}}
\put(20,20){\vector(-1,-2){7}} \put(15,10){\line(-1,-2){5}}
\put(9.8,2){\OVLwa{4} \put(0,-2){\Vwa} }
\end{picture} }
\newcommand{\LCqc}{
\begin{picture}(20,20)(0,5) \linethickness{.3pt}
\put(0,20){\vector(1,-1){7}} \put(0,20){\line(1,-1){10}}
\put(10,10){\vector(1,-1){7}} \put(10,10){\line(1,-1){10}}
\put(10,-.2){\OVLwa{2}}
\end{picture} }
\newcommand{\LCQc}{
\begin{picture}(20,20)(0,5) \linethickness{.3pt}
\put(0,20){\vector(1,-1){7}} \put(0,20){\line(1,-1){10}}
\put(10,10){\vector(1,-1){7}} \put(10,10){\line(1,-1){10}}
\put(10,10){\OVLwa{2}}
\end{picture} }
\title{Quantum bundles and quantum interactions}
\date{{\small Revised version, 15 February 2006} }
\author{Daniel Canarutto\\[6pt]
{\small\it Dipartimento di Matematica Applicata ``G. Sansone'', }\\
{\small\it Via S. Marta 3, 50139 Firenze, Italia}\\
{\small email:~daniel.canarutto@unifi.it}\\
{\small http://www.dma.unifi.it/\char126 canarutto}}
\begin{document}
\bibliographystyle{alpha}
\maketitle
\begin{abstract}\noindent
A geometric framework for describing quantum particles
on a possibly curved background is proposed.
Natural constructions on certain distributional bundles (`quantum bundles')
over the spacetime manifold yield a quantum ``formalism'' along any
1-dimensional timelike submanifold (a `detector');
in the flat, inertial case this turns out to reproduce
the basic results of the usual quantum field theory,
while in general it could be seen as a local,
``linearized'' description of the actual physics.
\end{abstract}


\section*{Introduction}
Quantisation,
intended as the construction of a quantum theory
by applying suitable rules to classical systems,
is perhaps the most common approach
to the study of the foundations of quantum physics;
indeed, this philosophy
has produced an immense physical and mathematical literature.
There is, however, a widespread opinion that the true relation
between classical and quantum theories should rather go in the opposite sense:
at least in principle,
classical physics should derive from quantum physics,
thought to be more fundamental.

As a first step in that direction,
one could try and build a stand-alone mathematical model,
not derived from a quantisation procedure,
which should reproduce (at least) the basic observed facts
of elementary particle physics.
The present article is a proposal in this sense,
based on two main ingredients:
\emph{free states}, and \emph{interaction}.
A further interesting feature of the model is its freedom
from the requirement of spacetime flatness.

The fundamental mathematical tool of my exploration
is the geometry of \emph{distributional bundles},
that is bundles over classical (finite-dimensional Hausdorff) manifolds
whose fibres are distributional spaces.
These arise naturally from a class of finite-dimensional 2-fibred bundles,
which turns out to contain the most relevant physical cases.
The basics of their geometry have been exposed
in two previous papers~\cite{C00a,C04a}
along the line of thought stemming from Fr\"olicher's notion
of smoothness~\cite{Fr,FK,KM,MK,CK95}.

While I do not quantise classical fields, at this stage
I do consider certain finite-dimensional geometric structures
which are related to classical field theories.\footnote{
In particular gravitation, here, is a fixed background.}
From these one can naturally build 2-fibred bundles and, eventually,
\emph{quantum bundles}:
distributional bundles whose fibres are spaces of one-particle states,
and the related \emph{Fock bundles}.
It turns out that the underlying,
finite-dimensional geometric structure determines
a distinguished connection on a quantum bundle;
this connection is related to the description of \emph{free-particle states}.

The basic idea about quantum interactions is that they should be described
by a new connection on the Fock bundle,
obtained by adding an interaction morphism to the free-particle connection.
This approach requires the notion of a \emph{detector},
defined to be a timelike 1-dimensional submanifold of the spacetime manifold.
Then a natural interaction morphism indeed exists in the fibres
of the restricted Fock bundle.
It turns out that a detector carries a quantum ``formalism''
which can be seen as a kind of complicated clock;
in the flat, inertial case this turns out to reproduce
the basic results of the usual quantum field theory,\footnote{
The usual quantum fields can be recovered~\cite{C04b}
as certain natural geometric structures of the quantum bundles,
but they only play a marginal role in this approach.}
while in general it could be seen as a local, ``linearized'' description
of the actual physics.

The paper's plan is as follows.
In the two first sections I will summarize the basic ideas
about distributional bundles and quantum bundles,
the latter being defined as certain bundles of generalized half-densities
on classical momentum bundles;
then I will introduce generalized frames for quantum bundles and
the notion of a detector.
In section~\ref{S:Quantum interaction}
I will illustrate the construction of the quantum interaction
from a general (and necessarily sketchy) point of view.
In section~\ref{S:Scalar particles} these ideas will be implemented
in the simplest case, a theory of two scalar particles;
in sections~\ref{S:Electron and positron free states},
\ref{S:Photon free states},
\ref{S:Electromagnetic interaction}
and~\ref{S:QED} I will show how to treat QED in the above said setting;
in the flat inertial case one then recovers the basic known results.
Here, the role of 2-fibred bundles turns out to be specially meaningful.
\section{Distributional bundles}\label{S:Distributional bundles}
For details about the ideas reviewed in this section,
see~\cite{C00a,C04a}.\par
Let $\pp:\Y\to\ul\Y$ be a real or complex classical vector bundle,
namely a finite-dimensional vector bundle
over the Hausdorff paracompact smooth real manifold $\ul\Y$\,.
Moreover assume that $\ul\Y$ is oriented, let  $n:=\dim\ul\Y$\,,
and consider the positive component
$\VV^*\ul\Y:=(\weu{n}\TS\ul\Y)^+\to\ul\Y$\,,
called the bundle of \emph{positive densities} on $\ul\Y$\,.

Let $\YCo\equiv\DCo(\ul\Y\,,\VV^*\ul\Y\ten{\ul\Y}\Y^*)$ be the vector space
of all `test sections',
namely smooth sections $\ul\Y\to\VV^*\ul\Y\ten{\ul\Y}\Y^*$
which have compact support.
A topology on this space can be introduced by a standard procedure~\cite{Sc};
its topological dual will be denoted as
$\YC\equiv\DC(\ul\Y\,,\Y)$ and called the space of \emph{generalized sections},
or \emph{distribution-sections} of the given classical bundle.
Some particular cases of generalized sections are that of \emph{$r$-currents}
($\Y\equiv\weu{r}\TO^*\ul\Y$\,, $r\in\NN$)
and that of \emph{half-densities}
($\Y\equiv(\VV^*\ul\Y)^{1/2}\equiv\VV^{-1/2}\ul\Y$).

A curve $\a:\RR\to\YC$ is said to be \emph{F-smooth} if the map
$$\bang{\a,u}:\RR\to\CC:t\mapsto\bang{\a(t),u}$$
is smooth for every $u\in\YCo$\,.
Accordingly, a function $\phi:\YC\to\CC$ is called F-smooth
if $\phi\comp\a:\RR\to\CC$ is smooth for all F-smooth curve $\a$\,.
The general notion of F-smoothness,
for any mapping involving distributional spaces,
is introduced in terms of the standard smoothness of all maps,
between finite-dimensional manifolds,
which can be defined through compositions with F-smooth curves and functions.
Moreover, it can be proved that a function $f:\M\to\RR$\,,
where $\M$ is a classical manifold,
is smooth (in the standard sense) iff the composition $f\comp c$
is a smooth function of one variable for any smooth curve $c:\RR\to\M$\,.
Thus one has a unique notion of smoothness based on smooth curves,
including both classical manifolds and distributional spaces.

In the basic classical geometric setting underlying distributional bundles
one considers a classical $2$-fibred bundle
$$\begin{CD}\V @>{\displaystyle\qq}>>\E 
@>{\displaystyle\ul\qq}>> \B~, \end{CD}$$
where $\qq:\V\to\E$ is a vector bundle,
and the fibres of the bundle $\E\to\B$ are smoothly oriented.
Moreover, one assumes that $\qq\comp\ul\qq:\V\to\B$ is also a bundle,
and that for any sufficiently small open subset $\X\subset\B$
there are bundle trivializations
$$(\ul\qq\,,\ul\yy):\E\!_\Xx\to\X\times\ul\Y~,\quad
(\qq\comp\ul\qq\,,\yy):\V\!\!_\Xx\to\X\times\Y$$
with the following projectability property:
there exists a surjective submersion $\pp:\Y\to\ul\Y$
such that the diagram
$$\begin{CD}
\V\!\!_\Xx @>{\displaystyle(\qq\comp\ul\qq\,,\yy)}>> \X\times\Y \\
@V{\displaystyle\qq}VV @VV{\displaystyle\Id{\Xx}\times\pp}V \\
\E\!_\Xx @>>{\displaystyle(\ul\qq,\ul\yy)}> \X\times\ul\Y
\end{CD}$$
commutes; this implies that $\Y\to\ul\Y$ is a vector bundle,
which is not trivial in general.

The above conditions are easily checked to hold in many cases
which are relevant for physical applications,
and in particular when $\V=\E\cart{\B}\W$ where $\W\to\B$ is a vector bundle,
when $\V=\VO\!\E$ (the vertical bundle of $\E\to\B$)
and when $\V$ is any component of the tensor algebra of $\VO\!\E\to\E$\,.

For each $x\in\B$ one considers the distributional space
$\VC\!_x:=\DC(\E_x\,,\,\V\!_x)$, and obtains the fibred set
$$\wp:\VC\equiv\DC\!_\Bb(\E,\V):=\bigsqcup_{x\in\B}\,\VC\!_x\to\B~.$$
An isomorphism of vector bundles yields an isomorphism of the corresponding
spaces of generalized sections;
hence, a local trivialization
of the underlying classical 2-fibred bundle, as above,
yields a local bundle trivialization
$$(\wp,\Ysf):\VC\!\!_\Xx\to\X\times\YC~,\quad
\YC\equiv\DC(\ul\Y\,,\Y)$$   
of $\VC\to\B$\,.
Moreover, a smooth atlas of 2-bundle trivializations
determines a linear F-smooth bundle atlas on $\VC\to\B$\,,
which is said to be an \emph{F-smooth distributional bundle}.
In general, the F-smoothness of any map from or to $\VC$
is equivalent to the F-smoothness of its local trivialized expression.

One defines the \emph{tangent space} of any F-smooth space
through equivalence classes of F-smooth curves;
tangent prolongations of any F-smooth mappings can also be shown to exist.
Thus one gets, in particular, the tangent space $\TO\VC$,
which has local trivializations as $\TO\X\times\TO\YC$,
its \emph{vertical subspace} and the \emph{first jet bundle} $\JO\VC\to\VC$.
A \emph{connection} is defined to be an F-smooth section $\Gfr:\VC\to\JO\VC$.

With some care, many of the usual chart expressions
of finite-dimensional differential geometry can be extended
to the distributional case.
In particular, let $\s:\B\to\VC$ be an F-smooth section
and $\s^\yYsf:=\Ysf\comp\s:\B\to\YC$ its `chart expression'.
Then its covariant derivative has the chart expression
$$(\nabla\s)^\yYsf
=\dot\xx^a\,(\de_a\s^\yYsf-\Gfr_{\yYsf a}\,\s^\yYsf)~,$$
where $(\xx^a)$ is a chart on $\X\subset\B$
and $\Gfr_{\yYsf a}:\X\to\End(\YC)$\,,~$a=1,\dots,\dim\B$\,.

The notions of \emph{curvature} and of \emph{adjoint connection}
can also be introduced.
Furthermore, it can be shown that any projectable connection
on the underlying classical 2-bundle determines a distributional connection;
however, not all distributional connections arise from classical ones.
\section{Quantum bundles}\label{S:Quantum bundles}
Let $\LL$ be the semi-vector space of \emph{length units}
(see \cite{CJM,C00b} for a review of unit spaces) and $(\M,g)$ a spacetime.
The spacetime metric $g$ has `conformal weight' $\LL^2\cong\LL\tn\LL$\,,
namely it is a bilinear map $\TO\M\cart{\M}\TO\M\to\LL^2$,
while its inverse $g^\#$ has conformal weight $\LL^{-2}\cong\LL^*\tn\LL^*$.

For $m\in\LL^{-1}\cong\LL^*$ let $\Pm\cong\K_m^+\subset\TS\M$
be the subbundle over $\M$ of all future-pointing $p\in\TS\M$
such that\footnote{Throughout this paper,
the signature of the metric is $(1,3)$\,;
moreover $\hbar=c=1$\,, so that $\LL$ is the unique unit space involved.}
$g^\#(p,p)=m^2$\,.
Then $\Pm$ is the classical \emph{phase bundle} for a particle of mass $m$\,;
the case $m=0$ can be also considered.
Furthermore, consider the 2-fibred bundle
$$\bigl[(\weu3\TS\Pm)^+\bigr]{}^{1/2}\equiv\VV^{-1/2}\Pm\to\Pm\to\M$$
whose upper fibres are the spaces of half-densities
on the fibres of $\Pm\to\M$.
There is a distinguished section
$$\sqrt{\om_m}:\Pm\to\LL^{-1}\tn\VV^{-1/2}\Pm~;$$
here, $\om_m$ is the \emph{Leray form} of the hyperboloids
(the fibres of $\Pm\to\M$),
usually indicated as $\d(\gf-m^2)$ where $\gf$ is the contravariant
quadratic form associated with the metric.
If $(\pp_\l)=(\pp_0,\pp_i)$ are $\LL^{-1}$-scaled orthonormal coordinates
on the fibres of $\TS\M\to\M$\,,
then one finds the coordinate expression
$$\sqrt{\om_m}=\frac{\sqrt{\dO^3\ps}}{\sqrt{2\,\Eo_m}}~,$$
where
$\Eo_m:=\sqrt{m^2+|\ps^2|}=\sqrt{m^2+\d^{ij}\pp_i\,\pp_j}$
is indicated simply as $\pp_0$ if no confusion arises.
Note that
$\dO^3\ps\equiv\dO\pp_1\we\dO\pp_2\we\dO\pp_3$
is the ``spatial'' (scaled)
volume form determined by the ``observer'' associated with the coordinates,
ans can be seen as a volume form on the fibres of $\Pm$
via orthogonal projection.

It can be seen~\cite{C04a} that the spacetime connection $\G$
determines a connection $\G_{\!\!m}$ of $\Pm\to\M$\,,
as well as a linear connection of the 2-fibred bundle $\VS\Pm\to\Pm\to\M$
which is projectable on $\G_{\!\!m}$\,;
on turn this determines a linear projectable connection $\hat\G_{\!\!m}$
of $\VV^{-1/2}\Pm\to\Pm\to\M$\,,
with the coordinate expression
$$(\G_{\!\!m})_{ai}=-\G\iIi a0i\,\pp_0-\G\iIi aji\,\pp_j~,\quad
(\hat\G_{\!\!m})_a=-\frac{\G\iIi a0i\,g^{ij}\,\pp_j}{\pp_0}+\oh\,\G\iIi aii$$
(here $a$ is an index for coordinates on $\M$ and the spacelike
coordinates $(\pp_j)$ play the role of fibre coordinates on $\Pm\to\M$).

Next consider the distributional bundle
$$\PC_{\!\!m}:=\DCh\!_\Mm(\Pm)
\equiv\DC\!_\Mm(\Pm,\CC\tn\VV^{-1/2}\Pm)\to\M~,$$
whose fibre over each $x\in\M$ is the vector space of all
(complex-valued) generalized half-densities on $(\Pm)_x$\,.
The connection $\hat\G_{\!\!m}$ determines a smooth
(in Fr\"olicher's sense) connection $\PC_{\!\!m}\to\M$
which can be characterized in various ways~\cite{C04a},
the most simple being the following:
let $c:\RR\to\M$ be any local curve and $p:\M\to\Pm$ a local section
which is parallely transported along $c$\,;
then the local section
$$\d_p\tn(\om_m)^{-1/2}:\M\to\LL\tn\PC_{\!\!m}
:x\mapsto\d_{p(x)}\tn(\om_m)^{-1/2}~,\quad x\in\M$$
is parallely transported along $c$\,,
where $\d_{p(x)}$ denotes the Dirac density on $(\Pm)_x$
whose support is the point $p(x)$\,.

Let now $\V\to\Pm\to\M$ be a (real or complex) 2-fibred vector bundle,
and consider the distributional bundle
$$\VC^1:=\DCh\!_\Mm(\Pm,\V)
\equiv\DC\!_\Mm(\Pm,\VV^{-1/2}\Pm\ten{\Pm}\V)\to\M~,$$
whose fibre over each $x\in\M$ is the vector space of all $\V$-valued
generalized half-densities on $(\Pm)_x$\,.
In practice, this $\V$ will be related to the bundle whose sections
are the fields of the classical theory which, in the usual approach,
correspond to the quantum theory under consideration.
One could think that it suffices to deal
with a ``semi-trivial'' 2-fibred bundle
$\Pm\cart{\M}\V$ where $\V\to\M$ is a vector bundle,
however it will be seen (\S\ref{S:QED}) that the general setting
is actually needed.

\remark~
If a Hermitian metric on the fibres of $\V$ is given,
then one can define a \emph{Hilbert bundle} $\HC\to\M$\,,
and has inclusions $\VC^1_{\!{\circ}}\subset\HC\subset\VC^1$
(where $\VC^1_{\!{\circ}}\to\M$ is the subbundle whose fibres
are constituted by test sections);
namely one has a bundle of `rigged Hilbert spaces'~\cite{BLT}.
\smallbreak

A \emph{Fock bundle} can be constructed as
$$\VC:=\bigoplus_{j=0}^\infty\VC^j~,$$
where
$$\text{either}\quad \VC^j:=\weu{j}\VC^1\qquad \text{or}\quad
\quad \VC^j:=\vee^j\VC^1$$
(antisymmetrized and symmetrized tensor products).

If a connection $\g$ of $\V\to\Pm\to\M$
linear projectable over $\G_{\!\!m}$ is given
(which is the case in most physical situations),
then one also gets a connection of $\VC^1\to\M$\,;
this can be naturally extended to a connection on $\VC\to\M$,
which will be called the \emph{free particle connection}.
For any local section $\s:\M\to\VC^1$ one has the coordinate expression
$$\Cfr_a^\sA(\s)=\g\iIi a\sA\sB\,\s^\sB-(\G_{\!\!m})_{ai}\,\de^i\s^\sA~.$$
\section{Generalized frames}\label{S:Generalized frames}
For each $p\in(\Pm)_x$\,, $x\in\M$,
let $\d[p]$ denote the Dirac generalized density on the fibre $(\Pm)_x$
with support $\{p\}$\,;
namely, if $f:(\Pm)_x\to\CC$ is a test function then one has
$\bang{\d[p]\,,f}=f(p)$\,.
It can be written as\footnote{
While $p$ denotes an element of $\Pm$\,,
the sans-serif symbol $\pp$ is used for the fibre coordinates.}
$$\d[p]=\br\d[p]\,\dO^3\ps$$
where $\br\d[p]$ is the $\LL^3$-scaled \emph{generalized function},
usually denoted as $\br\d[p](q)\equiv\d(q-p)$\,,
acting on test densities $\phi=\br\phi\,\dO^3\ps$ as
$$\bang{\br\d[p],\phi}=\bang{\d[p],\br\phi}=\br\phi(p)~.$$
Actually any generalized density can be expressed in this way
as a generalized function times a given volume form;
moreover, note that the spacelike volume form $\dO^3\ps$\,,
as well as the induced volume form on the fibres of $\Pm\to\M$
denoted in the same way,
only depends on the choice of an `observer'
(\ie\ a timelike future-pointing unit vector field)
and not on the particular frame of $\TS\M$ adapted to it.

Let now $l\in\LL$ be an arbitrarily fixed length unit,
and consider the unscaled generalized half-density
$$\Bsf_p:=l^{-3/2}\,\br\d[p]\,\sqrt{\dO^3\ps}
=\frac1{\sqrt{2\,l^3\,\pp_0}}\,\d[p]\tn\om_m^{-1/2}~,$$
acting on test half-densities $\theta=\br\theta\,\sqrt{\dO^3\ps}$ as
$$\bang{\Bsf_p\,,\theta}=l^{-3/2}\,\bang{\d[p],\br\theta}
=l^{-3/2}\,\br\theta(p)~;$$
for each $x\in\M$ the set $\{\Bsf_p\}$\,, $p\in(\Pm)_x$\,,
can be seen as a \emph{generalized frame}
of the distributional bundle $\PC_{\!\!m}$ at $x$\,.
Let moreover $\{\bb_\sA\}$ be a frame
of the classical vector bundle $\V\to\Pm$\,, ${\scriptstyle A}=1,...,n$\,; then
$$\{\Bsf_{p\sA}\}:=\{\Bsf_p\tn\bb_\sA\}$$
is a generalized frame of $\VC^1\to\M$\,.
In fact, any $\psi=\psi^\sA\,\bb_\sA\in\VC^1$ can be written as
$\int\psi^\sA(\pp)\,\Bsf_{\pp\sA}$\,,
which is to be intended in the generalized sense
$$\bang{\psi,\theta}=
\int \br\psi^\sA(\pp)\,\br\theta_{\sA}(\pp)\,\dO^3\pp_\sbo$$
where $\theta\in\VC^1_{\!{\circ}}$ is a test half density
in the same fibre as $\psi$\,.

Let $\AC$ be a set (index set);
a \emph{generalized multi-index} is defined to be a map
$$I:\AC\to\{0\}\cup\NN$$
vanishing outside some finite subset $\AC_I\subset\IC$\,;
it can be represented through its graphic
$$\bigl\{ (\a_1\,,I_1),(\a_2\,,I_2),\dots,(\a_r\,,I_r)\bigr\}~,
\quad \AC_I=\{\a_1\,,\dots\,,\a_r\}$$
for any (arbitrary and inessential) ordering of $\AC_I$\,.
Now one extends the generalized frame $\{\Bsf_{p\sA}\}$
to a generalized frame of the Fock bundle $\VC\to\M$
by letting $\AC_x=(\Pm)_x\,{\times}\,\{1,...,n\}$ for each $x\in\M$\,,
and setting
\begin{align*}
&\Bsf_I:=\frac{
(\Bsf_{\a_1})^{I_1}\ve(\Bsf_{\a_2})^{I_2}\ve\cdots(\Bsf_{\a_r})^{I_r}
}{\sqrt{I_1!\,I_2!\,\cdots\,I_r!}}
&& \text{(bosons)},
\\[6pt]&
\Bsf_I:=\Bsf_{\a_1}\we\Bsf_{\a_2}\we\cdots\we\Bsf_{\a_r}
&& \text{(fermions)},
\end{align*}
where
$$(\Bsf_\a)^k:=\underbrace{\Bsf_\a\ve\cdots\ve\Bsf_\a}_{k~\text{times}}~.$$
In a more detailed way one writes $\a_i=(p_i\,,{\scriptstyle A}_i)$ and
\begin{align*}&
\Bsf_I:=\frac{
(\Bsf_{p_1\sA_1})^{I_1}\ve(\Bsf_{p_2\sA_2})^{I_2}
\ve\cdots(\Bsf_{p_r\sA_r})^{I_r}}{\sqrt{I_1!\,I_2!\,\cdots\,I_r!}}
&& \text{(bosons)},
\\[6pt]&
\Bsf_I:=\Bsf_{p_1\sA_1}\we\Bsf_{p_2\sA_2}\we\cdots\we\Bsf_{p_r\sA_r}
&& \text{(fermions)}.
\end{align*}
If one has a Hermitian structure in the fibres of $\V\to\Pm$
and $\{\bb_\sA\}$ is an orthonormal classical frame,
then one gets an `orthonormality' relation $\bang{\Bsf_I\,,\Bsf_J}=\d_{IJ}$\,,
to be interpreted in a generalized (\ie\ distributional) sense.
\section{Detectors}\label{S:Detectors}
By a `detector' I mean a 1-dimensional time-like submanifold $\T\subset\M$\,.
Locally this determines, via the exponentiation map, a \tps\ splitting,
which in a sense relates the momentum-space based approach presented here
to a position-space approach,
though the relation is precise only if the induced splitting is global.

Consider restrictions of the quantum bundles previously introduced
to bundles over $\T$, so write
$$\PC_{\!\!m}\to\T~,\quad \VC^1\to\T~,\quad \VC\to\T~,$$
and the like.
Clearly, the free particle connection determines connections of these bundles;
it actually turns out that one gets (possibly local) splittings of them.
So one writes, for example
$$\VC\cong\T\times\VC_{\!t_0}$$
where $t_0\in\T$ is some arbitrarily chosen point.
Note that the free particle connection, by construction,
preserves ``particle number'',
namely is reducible to a connection of each of the subbundles $\VC^j$,
$j\in\{0\}\cup\NN$\,.

The unit future-pointing vector field $\Theta_0:\T\to\LL\tn(\TS\M)_\Tt$
tangent to $\T$ determines an orthogonal splitting
$(\TS\M)_\Tt=\TS\T\cart{\T}(\TS\M)_\Tt^\sbo$,
and a diffeomorphism $(\Pm)_\Tt\leftrightarrow(\TS\M)_\Tt^\sbo$\,;
the `spacelike' volume form on the fibres of $\LL\tn(\TS\M)_\Tt^\sbo$
then yields a scaled volume form on the fibres of $(\Pm)_\Tt$\,;
with the choice of a length unit
one obtains a generalized frame $\{\Bsf_{p\sA}\}$ of $\VC^1\to\T$\,;
in practice, this is defined in the same way as
the generalized frame of $\VC^1\to\M$
introduced in~\S\ref{S:Generalized frames},
where now the orthonormal coordinates $(\pp_\l)\equiv(\pp_0\,,\pp_i)$
are \emph{adapted} to the above said splitting.

Let now $p:\T\to(\Pm)_\Tt$ be a covariantly constant section
and $\bigl(\bb_\sA(p)\bigr)$ a frame of $\V\to\Pm$
covariantly constant over $p$\,.
If $\T\subset\M$ is a geodesic submanifold,
then $\Bsf_{p\sA}$ is covariantly constant along $\T$
relatively to the free-particle connection.
Thus the generalized orthonormal set $\{\Bsf_{p\sA}\}$
indexed by covariantly constant sections $p:\T\to(\Pm)_\Tt$
and by the classical index ${\scriptstyle A}$ is constant
in the same sense.
If $\T$ is not geodesic then one can either construct the generalized frame
at some chosen $t_0\in\T$ and then parallely propagate it along $\T$,
or modify the definition of the free-particle connection of $\VC\to\T$
by relating it to \emph{Fermi transport}
rather than parallel transport along $\T$.
From the physical point of view one may expect different interpretations
of these two settings,
which however give rise essentially to the same formalism.
\section{Quantum interaction}\label{S:Quantum interaction}
The general idea of quantum interaction is the following.
Consider a Fock bundle
$\VC=\VC'\ten{\T}\VC''\ten{\T}\VC'''\ten{\T}\dots$
(each factor being, on turn, a Fock bundle accounting for a given particle type)
endowed with a free-particle connection $\Gfr$\,.
Suppose that there exists a distinguished section
$\Hfr:\T\to\LL^{-1}\tn\End(\VC)$\,;
by considering the unit future-oriented section
$\dt:\T\to\LL\tn\TS\T$ determined via the spacetime metric,
one can introduce a new connection $\Gfr{-}\iO\,\Hfr\,\dt$
(possibly mixing particle numbers and types).
A section $\T\to\VC$ which is constant relatively to $\Gfr{-}\iO\,\Hfr\,\dt$
describes the evolution of a particle system, or rather
the evolution of a quantum \emph{clock} (in a broad sense) of the detector.
This evolution can be compared to that determined by $\Gfr$ alone,
namely it can be read in a fixed Fock \emph{space} $\YC\equiv\VC_{t_0}$\,,
$t_0\in\T$,
such that $\VC\cong\T\times\YC$ is the splitting
determined by $\Gfr$\,.\footnote{Related ideas,
describing the evolution of a quantum system in terms of a connection
on a functional bundle in a Galileian setting,
have been introduced in~\cite{JM,CJM}.}

Now the evolution operator $\Ucal_{t_0}:\T\to\End(\YC)$
can be written as the formal series
$$\Ucal_{t_0}(t)=\id+\sum_{N=1}^\infty \frac{(-\iO)^N}{N!}\,
\int_{t_0}^t\dt_1 \int_{t_0}^t\dt_2 \dots \int_{t_0}^t\dt_N\,
\topf{\Hfr(\tt_1)\,\Hfr(\tt_2)\dots\Hfr(\tt_N)}~,$$
where $\topf{\cdot\cdot}$ denotes the \emph{time-ordered product};
the \emph{scattering operator} is defined to be
$\Scal:=\Ucal_{-\infty}(+\infty)\in\End(\YC)$\,.
However, besides any convergence questions,
the basic problem is the existence of $\Hfr$\,;
actually I'm going to show that there is a natural way of introducing it,
and a way which is consistent with the results of the standard theory,
but only as a morphism $\VCo\to\VC$
(where $\VCo\subset\VC$ is the subbundle of test elements).
This implies that many \emph{single terms} of the above series
are not defined.
Nevertheless, parts of it give considerable information
which turns out to be physically true,
at least in the standard, flat spacetime situation.
Furthermore, in some way $\Scal$ turns out to be well-defined
in renormalizable theories.

In the rest of this section I will expose the basic ideas
for the construction of $\Hfr$\,.

For each $m\in\{0\}\cup\LL^{-1}$ the spacetime geometry
yields $\LL^{-3}$-scaled volume forms $\eta_m=\dO^3\ps$
on the fibres of $(\TS\M)^\sbo\to\T$\,,
giving rise to equally scaled volume forms,
denoted by the same symbols, on the fibres of $\Pm\to\T$.

Now for $m',m'',m'''\in\{0\}\cup\LL^{-1}$ consider the bundle `of three momenta'
$$\Ptru:=\Px{m'}\cart{\M}\Px{m''}\cart{\M}\Px{m'''}\to\M~,$$
and the section of \emph{scaled} densities
$$\d\!_{\sst\vartriangle}:\T\to\LL^{-6}\tn\DC(\Ptru)
=\LL^{-6}\tn\DC(\Px{m'})\ten{\T}\DC(\Px{m''})\ten{\T}\DC(\Px{m'''})~,$$
given by
\begin{align*}
\bang{\d\!_{\sst\vartriangle},f}&:=
\iint f(\pp'_\sbo\,,\pp''_\sbo\,,
-\pp'_\sbo{-}\pp''_\sbo)\,\eta_{m'}\we\eta_{m''}
=\iint f(\pp'_\sbo\,,\pp''_\sbo\,,-\pp'_\sbo{-}\pp''_\sbo)\,
\dO^3\pp'\,\dO^3\pp''
\\[6pt]
&\phantom:=
\iint f(\pp'_\sbo\,,\pp''_\sbo\,,\pp'''_\sbo)\,
\d(\pp'_\sbo{+}\pp''_\sbo{+}\pp'''_\sbo)\,\dO^3\pp'\,\dO^3\pp''\,\dO^3\pp'''~,
\end{align*}
where $f\in\DCo(\Ptru)$\,.
It can be writen in the form
$$\d\!_{\sst\vartriangle}=\br\d\,\eta_{m'}\tn\eta_{m''}\tn\eta_{m'''}
=\br\d\!_{\sst\vartriangle}\,\dO^3\pp'\tn\dO^3\pp''\tn\dO^3\pp'''~,$$
where
$\br\d\!_{\sst\vartriangle}\equiv\d(\pp'_\sbo{+}\pp''_\sbo{+}\pp'''_\sbo)$
is an $\LL^3$-valued generalized function.

Now one introduces the true (unscaled) generalized half-density
$$\ul\Lambda:=\br\d\!_{\sst\vartriangle}\,
\sqrt{\om_{m'}}\tn\sqrt{\om_{m''}}\tn\sqrt{\om_{m'''}}
:\T\to\DCh(\Ptru)
=\DCh(\Px{m'})\ten{\T}\DCh(\Px{m''})\ten{\T}\DCh(\Px{m'''})~,$$
which has the coordinate expression
$$\ul\Lambda=\frac{\d(\pp'_\sbo{+}\pp''_\sbo{+}\pp'''_\sbo)}%
{\sqrt{8\,\pp_0'\pp_0''\pp_0'''}}\,
\sqrt{\dO^3\pp'}\tn\sqrt{\dO^3\pp''}\tn\sqrt{\dO^3\pp'''}~.$$
The fact that $\ul\Lambda$ is unscaled,
independently of the choice of a length unit,
will turn out to be essential for its role in the quantum interaction;
here it will describe the interaction of three particles,
but clearly it can be readily generalized
for describing the interaction of any number of particles.
The different particle types are characterized by different
complex 2-fibred bundles $\V'\to\Px{m'}\to\M$ and the like,
and one must have a `classical interaction Lagrangian'
that is a scalar-valued 3-linear contraction among the fibres;
this is a section
$$\ell\interaction:\Ptru\to
\V'{}^\lin\ten{\Ptru}\V''{}^\lin\ten{\Ptru}\V'''{}^\lin~,$$
which can be seen as 3-linear fibred contraction.
The structure of these bundles must allow for
`index raising and lowering',
thus yielding a number of objects related to $\ell\interaction$
and distinguished by various combinations of index types.
Of course these arise in the easiest way when one has
fibred Hermitian structures of the considered bundles
(the fundamental case of electrodynamics, however,
will be seen [\S\ref{S:QED}] to be somewhat more involved).
In particular,
$\ell\interaction^\dag:\Ptru\to\V'\ten{\Ptru}\V''{}\ten{\Ptru}\V'''$\,.

Now one gets a section
$$\Lambda:=\ul\Lambda\tn\ell\interaction^\dag:
\T\to\DCh_\Tt(\Ptru\,,\V'\ten{\Ptru}\V''\ten{\Ptru}\V''')
\equiv\VC'{}^1\ten{\T}\VC''{}^1\ten{\T}\VC'''{}^1~,$$
where $\VC'{}^1:=\DCh_\Tt(\Ptru\,,\V')$ and the like.
The essential idea of the quantum interaction is then the following:
make $\Lambda$ act in the fibres of the Fock bundle
$\VC\equiv\VC'\ten{\T}\VC''\ten{\T}\VC'''\to\T$
by using each one of its tensor factors either as `absorption' (contraction)
or as `creation' (tensor product).
However, a fundamental issue is immediately apparent
(and will be furtherly discussed later on):
in general, this action is only well-defined on the subbundle
$\VCo\subset\VC$ of test elements,
so actually it gives rise to a morphism $\VCo\to\VC$
whose extendibility will have to be carefully examined.

The various `index types' of $\ell\interaction$ correspond to
the various actions performed by the corresponding tensor factors:
a covariant index determines a particle absorption,
a contravariant index determines a particle creation.
Furthermore one considers different types of $\ul\Lambda$\,,
each one to be coupled to a corresponding type of $\ell\interaction$
and obtained by changing the \emph{sign} of the momenta in the $\d$
generalized function.
So, for example, the type of $\ell\interaction$ which is a section
$\Ptru\to\V'{}^\lin\ten{\Ptru}\V''\ten{\Ptru}\V'''$
(the first factor is an absorption factor,
the second and third are creation factors)
is tensorialized by the `version' of $\ul\Lambda$ which has
$\d(-\pp'_\sbo{+}\pp''_\sbo{+}\pp'''_\sbo)$ in its coordinate expression.
In practice, I find it convenient using a `generalized index' notation
in which generalized indices are either high or low,
and repeated momentum indices are to interpreted as integration indices
(just as repeated ordinary indices are interpreted
as ordinary summation indices).
So, for
$f=\br f\,\sqrt{\dO^3\pp'}\tn\sqrt{\dO^3\pp''}\tn\sqrt{\dO^3\pp'''}\in\VCo$\,,
I'll write
\begin{align*}
& f=f_{\pp'\pp''\pp'''}\,\Bsf^{\pp'}\tn\Bsf^{\pp''}\tn\Bsf^{\pp'''}~,
&& f_{\pp'\pp''\pp'''}=l^{-9/2}\,\br f(\pp',\pp'',\pp''')~,
\\[6pt]
& \ul\Lambda
=\ul\Lambda^{\pp'\pp''\pp'''}\,\Bsf_{\pp'}\tn\Bsf_{\pp''}\tn\Bsf_{\pp'''}~,
&& \ul\Lambda^{\pp'\pp''\pp'''}=
\frac{\d(\pp'_\sbo{+}\pp''_\sbo{+}\pp'''_\sbo)}%
{\sqrt{8\,l^9\,\pp_0'\pp_0''\pp_0'''}}
\end{align*}
(where $\Bsf^{\pp}\equiv\Bsf_{\pp}$),
to be intended as
$$\bang{\ul\Lambda,f}=\ul\Lambda^{\pp'\pp''\pp'''}\,f_{\pp'\pp''\pp'''}=
\int\frac{\d(\pp'_\sbo{+}\pp''_\sbo{+}\pp'''_\sbo)}%
{\sqrt{8\,\pp_0'\pp_0''\pp_0'''}}\,\br f(\pp',\pp'',\pp''')\,
\dO^3\pp'\,\dO^3\pp''\,\dO^3\pp'''$$
(remark: the generalized index summation is performed via the unscaled
volume form $l^3\,\dO^3\ps$).

Analogously, the various index types of $\ul\Lambda$ (there are 8 of them)
can be written as
$$\ul\Lambda_{\pp'}{}^{\pp''\pp'''}\,
\Bsf^{\pp'}\tn\Bsf_{\pp''}\tn\Bsf_{\pp'''}~,
\quad
\ul\Lambda_{\pp'\pp''}{}^{\pp'''}\,\Bsf^{\pp'}\tn\Bsf^{\pp''}\tn\Bsf_{\pp'''}~,
~~\dots\quad\text{etcetera},$$
where
$$\ul\Lambda_{\pp'}{}^{\pp''\pp'''}
=\frac{\d(-\pp'_\sbo{+}\pp''_\sbo{+}\pp'''_\sbo)}%
{\sqrt{8\,\pp_0'\pp_0''\pp_0'''}}~,\quad
\ul\Lambda_{\pp'\pp''}{}^{\pp'''}
=\frac{\d(-\pp'_\sbo{-}\pp''_\sbo{+}\pp'''_\sbo)}%
{\sqrt{8\,\pp_0'\pp_0''\pp_0'''}}~,$$
and so on.
Correspondingly, the various types of $\Lambda$ can be written as
\begin{align*}
&\Lambda^{\pp'\sA',\pp''\sA'',\pp'''\sA'''}\,
\Bsf_{\pp'\sA'}\tn\Bsf_{\pp''\sA''}\tn\Bsf_{\pp'''\sA'''}~,&&
\Lambda^{\pp'\sA',\pp''\sA'',\pp'''\sA'''}=
\Lambda^{\pp'\pp''\pp'''}\,(\ell\interaction)^{\sA'\sA''\sA'''}~,
\\[6pt]
&\Lambda_{\pp'\sA'}{}^{\pp''\sA'',\pp'''\sA'''}\,
\Bsf^{\pp'\sA'}\tn\Bsf_{\pp''\sA''}\tn\Bsf_{\pp'''\sA'''}~,&&
\Lambda_{\pp'\sA'}{}^{\pp''\sA'',\pp'''\sA'''}=
\Lambda_{\pp'}{}^{\pp''\pp'''}\,(\ell\interaction)_{\sA'}{}^{\sA''\sA'''}~,
\\[-2pt]&\dots\dots&&\dots\dots\\[-2pt]
&\Lambda_{\pp'\sA',\pp''\sA'',\pp'''\sA'''}\,
\Bsf^{\pp'\sA'}\tn\Bsf^{\pp''\sA''}\tn\Bsf^{\pp'''\sA'''}~,&&
\Lambda_{\pp'\sA',\pp''\sA'',\pp'''\sA'''}=
\Lambda_{\pp'\pp''\pp'''}\,(\ell\interaction)_{\sA'\sA''\sA'''}~.
\end{align*}

The morphism $\Hfr:\VCo\to\LL^{-1}\tn\VC$
is essentially a sum whose terms are the various
types of $\Lambda$\,, with a further ingredient:
each term also has a factor
$$\l\,\eO^{-\iO\,(\pm\pp_0'\pm\pp_0''\pm\pp_0''')\,\tt}~,$$
where $\l\in\LL^{-1}$ is a constant;
the signs in the exponential match those of the corresponding spatial momenta.
Then $-\iO\,\Hfr\tn\dt:\T\to\VCo\tn\VC\tn\TS\T$
is the interaction term which modifies the free-particle connection.

The reader will note that, according to the setting above sketched,
the elements $\Bsf_{\pp\sA}$ and $\Bsf^{\pp\sA}$
in the various generalized frames
can be thought of, essentially, as the usual creation and absorption operators.
Moreover one could obtain further types of $\Lambda$ by exchanging tensor factors;
this only make a difference if particles of the same type are involved,
and is settled by considering only those terms
in which the creation operators stand on the right.

\section{Scalar particles}\label{S:Scalar particles}
Let's see how, in practice, the somewhat sketchy ideas exposed
in~\S\ref{S:Quantum interaction} can be implemented in the simplest case.
Many of the arguments used in this section are more or less standard,
the point is to show how they arise from a not-so-standard approach.

Consider a model of two  types of scalar particles,
one of mass $m$ and one massless,
with one-particle state bundles
$\VC'{}^1\equiv\PC_{\!\!m}$ and $\VC''{}^1\equiv\PC_{\!\!_0}$
and generalized frames $\{\Asf_p\}$\,, $p\in\Pm$\,,
and $\{\Bsf_k\}$\,, $k\in\Pz$\,.
The classical interaction is assumed to be just a constant
$\ell\in\LL^{-1}$,
and it incorporates the $\l$ introduced above.

At first-order, the formal series expression of the scattering operator is $\Scal=\id+\Scal_1$ with
$\Scal_1=-\iO\,\int_{-\infty}^{+\infty}\Hfr(\tt)\,\dt$\,.
In terms of generalized index notation,
one says that $\Scal_1$ has a `matrix element'
$$(\Scal_1)^\kk_{\pp\qq}
=\frac{-\iO\,\ell\,\d(-\ps+\ks-\qs)}{\sqrt{8\,l^9\,\kk_0\,\pp_0\,\qq_0}}
\int\limits_{-\infty}^{+\infty}\eO^{\iO\,(\kk_0-\pp_0-\qq_0)}\,\dt
=\frac{-2\pi\iO\,\ell\,\d(\kk-\pp-\qq)}{\sqrt{8\,l^9\,\kk_0\,\pp_0\,\qq_0}}~.$$
There are eight matrix elements of this kind,
each one describing \emph{one-point interaction}
and labelled by an elementary Feynman graph
(time running \emph{upwards}):
\begin{align*}
&\begin{picture}(20,20)(0,5)
\put(10,20){\line(-1,-2){10}}
\put(20,0){\line(-1,2){10}}
\linethickness{.2pt} \put(10,0){\OVLwa{4}} \put(10,18){\Vwa}
\end{picture}
&&(\Scal_1)_{\pp\qq\kk}=
\frac{-2\pi\iO\,\ell\,\d(-\pp-\kk-\qq)}{\sqrt{8\,l^9\,\pp_0\,\qq_0\,\kk_0}}~,
&&\qquad\begin{picture}(20,20)(0,5)
\put(10,10){\line(-1,1){10}}
\put(20,0){\line(-1,1){20}}
\linethickness{.2pt} \put(10,0){\OVLwa{2}}
\end{picture}
&&(\Scal_1)^\pp_{\qq\kk}=
\frac{-2\pi\iO\,\ell\,\d(\pp-\qq-\kk)}{\sqrt{8\,l^9\,\pp_0\,\qq_0\,\kk_0}}~,
\displaybreak[2]\\[8pt]
&\begin{picture}(20,20)(0,5)
\put(10,10){\line(-1,-1){10}}
\put(20,0){\line(-1,1){10}}
\linethickness{.2pt} \put(10,10){\OVRwa{2}}
\end{picture}
&&(\Scal_1)^\kk_{\pp\qq}=
\frac{-2\pi\iO\,\ell\,\d(\kk-\pp-\qq)}{\sqrt{8\,l^9\,\kk_0\,\pp_0\,\qq_0}}~,
&&\qquad\begin{picture}(20,20)(0,5)
\put(0,0){\line(1,1){20}}
\linethickness{.2pt} \put(10,0){\OVRwa{2}}
\end{picture}
&&(\Scal_1)_{\pp\kk}^\qq=
\frac{-2\pi\iO\,\ell\,\d(\qq-\pp-\kk)}{\sqrt{8\,l^9\,\kk_0\,\pp_0\,\qq_0}}~,
\displaybreak[2]\\[8pt]
&\begin{picture}(20,20)(0,5)
\put(20,0){\line(-1,1){20}}
\linethickness{.2pt} \put(10,10){\OVRwa{2}}
\end{picture}
&&(\Scal_1)^{\kk\pp}_\qq=
\frac{-2\pi\iO\,\ell\,\d(\kk+\pp-\qq)}{\sqrt{8\,l^9\,\kk_0\,\pp_0\,\qq_0}}~,
&&\qquad\begin{picture}(20,20)(0,5)
\put(10,10){\line(-1,1){10}} \put(20,20){\line(-1,-1){10}}
\linethickness{.2pt} \put(10,0){\OVRwa{2}}
\end{picture}
&&(\Scal_1)^{\pp\qq}_\kk=
\frac{-2\pi\iO\,\ell\,\d(\pp+\qq-\kk)}{\sqrt{8\,l^9\,\kk_0\,\pp_0\,\qq_0}}~,
\displaybreak[2]\\[8pt]
&\begin{picture}(20,20)(0,5)
\put(0,0){\line(1,1){20}}
\linethickness{.2pt} \put(10,10){\OVLwa{2}}
\end{picture}
&&(\Scal_1)_\pp^{\kk\qq}=
\frac{-2\pi\iO\,\ell\,\d(\kk+\qq-\pp)}{\sqrt{8\,l^9\,\kk_0\,\pp_0\,\qq_0}}~,
&&\qquad\begin{picture}(20,20)(0,5)
\put(10,0){\line(-1,2){10}}
\put(20,20){\line(-1,-2){10}}
\linethickness{.2pt} \put(10,0){\Vwa} \put(10,2){\OVRwa{4}}
\end{picture}
&&(\Scal_1)^{\pp\qq\kk}=
\frac{-2\pi\iO\,\ell\,\d(\pp+\qq+\kk)}{\sqrt{8\,l^9\,\kk_0\,\pp_0\,\qq_0}}~.
\end{align*}

Propagators arise when one consideris second-order matrix elements,
representing processes described, for example, by the diagrams
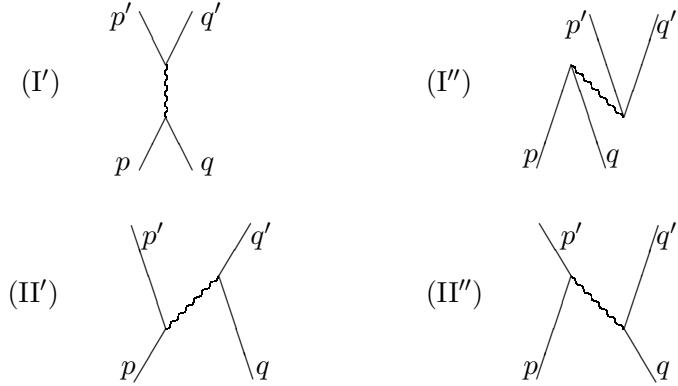
\begin{align*}
& \begin{picture}(40,60)(0,0)
\put(-35,30){(${\mathrm I}'$)}
\put(10,60){\line(1,-2){10}}
\put(20,40){\line(1,2){10}}
\put(10,0){\line(1,2){10}}
\put(20,20){\line(1,-2){10}}
\linethickness{.2pt} \put(20,20){\EVLwa{5}}
\put(1,0){$p$}
\put(-1,55){$p'$}
\put(33,0){$q$}
\put(33,55){$q'$}
\end{picture}
&&
\begin{picture}(60,60)(0,0)
\put(-35,30){(${\mathrm I}''$)}
\put(20,40){\line(-1,-3){13}} \put(20,40){\line(1,-3){13}}
\put(40,20){\line(-1,3){13}} \put(40,20){\line(1,3){13}}
\linethickness{.2pt} \put(39.5,20){\ODwRwa{6}}
\put(2,3){$p$}
\put(18,52){$p'$}
\put(33,3){$q$}
\put(52,52){$q'$}
\end{picture}
\displaybreak[2]\\[12pt]
&\begin{picture}(60,60)(0,0)
\put(-40,30){(${\mathrm{II}}'$)}
\put(20,20){\line(-1,3){13}} \put(20,20){\line(-3,-5){12}}
\put(40,40){\line(3,5){12}} \put(40,40){\line(1,-3){13}}
\linethickness{.2pt} \put(20,20){\ODeRwa{6}}
\put(3,3){$p$}
\put(11,52){$p'$}
\put(54,3){$q$}
\put(52,52){$q'$}
\end{picture}
&&
\begin{picture}(60,60)(0,0)
\put(-35,30){(${\mathrm{II}}''$)}
\put(20,40){\line(-3,5){12}} \put(20,40){\line(-1,-3){13}}
\put(40,20){\line(3,-5){12}} \put(40,20){\line(1,3){13}}
\linethickness{.2pt} \put(39.5,20){\ODwRwa{6}}
\put(2,3){$p$}
\put(16,52){$p'$}
\put(52,3){$q$}
\put(52,52){$q'$}
\end{picture}
\end{align*}
Here one has two types of second order processes
whose initial and final states contain two massive particles.
These types are labelled as (I) and (II),
and each of them comes in two subtypes,
distinguished by the time order of the two interactions involved
and respectively labelled as (I') and (I''), (II') and (II'').
By considering the form of the interaction,
one sees that the first diagram yields a contribution
\begin{align*}
(\Scal_{\mathrm I'})^{\pp'\!\qq'}_{\pp\qq}&=
\frac{-\ell^2}{l^6}
\int\limits_{-\infty}^{+\infty}\dt_2\int\limits_{-\infty}^{+\infty}\dt_1
\int\dO^3\kk\:\he(\tt_2-\tt_1)\,
\frac{\d(\ps'+\qs'-\ks)\,\d(\ks-\ps-\qs)}%
{\sqrt{16\,\pp_0'\,\qq_0'\,\pp_0\,\qq_0}\:2\,\kk_0}\cdot\\
&\hskip7.5cm
\cdot\eO^{\iO(-\pp_0+\kk_0-\qq_0)\tt_1}\,\eO^{\iO(\pp_0'-\kk_0+\qq_0')\tt_2}~,
\end{align*}
where $\he$ is the Heaviside function,
arising from the explicit expression of the time-ordered product
(which also yields two identical terms,
so the $1/2!$ factor in the $\Scal$ series cancels out).

Now one proceeds essentially in a more or less standard way.
First one uses a technical result: if $\vph$ is a test function
on any fibre of $\Pm\to\T$, then
\begin{align*}
\int\dO^3\ks\:\he(t)\,\eO^{\pm\iO\,\tt\,\Eo_m(\ks)}\,\vph(\ks)
&=\frac1{2\pi\iO}\,\lim_{\e\to0^+}
\int\dO^4\kk\:
\frac{\eO^{\iO\,\tt\,\kk_0}}{\kk_0\mp\Eo_m(\ks)-\iO\,\e}\:\vph(\ks)=
\\[6pt]
&=\frac1{2\pi\iO}\,\lim_{\e\to0^+}
\int\dO^4\kk\:
\frac{\eO^{-\iO\,\tt\,\kk_0}}{-\kk_0\mp\Eo_m(\ks)-\iO\,\e}\:\vph(\ks)
\end{align*}
(the proof uses the integral representation
$2\pi\iO\,\he(\tt)=\lim_{\e\to0^+}\int\limits_{-\infty}^{+\infty}
\frac{\eO^{\iO\,\tt\,\t}}{\t-\iO\,\e}\,\dO\t$
and some integration variable changes).
Eventually
$$(\Scal_{\mathrm I'})^{\pp'\!\qq'}_{\pp\qq}=
\frac{2\pi\iO\,\ell^2}{l^6\,\sqrt{16\,\pp_0'\,\qq_0'\,\pp_0\,\qq_0}}
\int\!\!\dO^4\kk\:
\frac{\d(\pp'+\qq'-\kk)\,\d(\kk-\pp-\qq)}{2\,|\ks|\,(-\kk_0+|\ks|-\iO\,\e)}$$
(the limit for $\e\to0^+$ is intended).
So, by a technical trick, an integral over $\T\times(\Pm)_{t_0}$\,,
$t_0\in\T$\,,
was transformed into an integral over a whole space $\TO^*_{\!\!t_0}\M$
of 4-momenta
(the momentum of the intermediate particle is `off-shell').

The calculation relative to the diagram $\mathrm{I}''$ is similar
but one has a few sign differences, getting
$$(\Scal_{\mathrm I''})^{\pp'\!\qq'}_{\pp\qq}=
\frac{2\pi\iO\,\ell^2}{l^6\,\sqrt{16\,\pp_0'\,\qq_0'\,\pp_0\,\qq_0}}
\int\!\!\dO^4\kk\:
\frac{\d(\pp'+\qq'-\kk)\,\d(\kk-\pp-\qq)}{2\,|\ks|\,(\kk_0+|\ks|-\iO\,\e)}~.$$
Finally,
$$(\Scal_{\mathrm I})^{\pp'\!\qq'}_{\pp\qq}=
(\Scal_{\mathrm I'})^{\pp'\!\qq'}_{\pp\qq}
+(\Scal_{\mathrm I''})^{\pp'\!\qq'}_{\pp\qq}=
\frac{-2\pi\iO\,\ell^2}{l^6\,\sqrt{16\,\pp_0'\,\qq_0'\,\pp_0\,\qq_0}}
\int\!\!\dO^4\kk\:
\frac{\d(\pp'+\qq'-\kk)\,\d(\kk-\pp-\qq)}{g(\kk,\kk)+\iO\,\e}~.$$

In case II one finds exactly the same result.

When the intermediate particle is massive one gets a similar expression,
with $g(\kk,\kk)$ in the propagator's expression
replaced by $g(\pp,\pp)-m^2$\,.

A word is due about the `infinities' arising when one considers diagrams
containing loops, as for example
$$
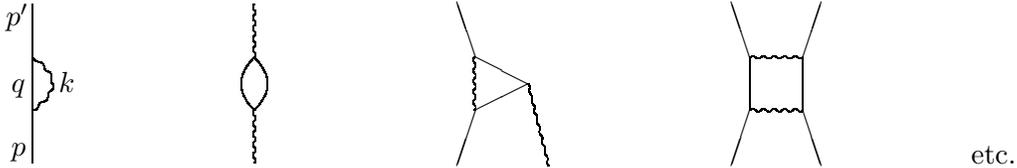
\begin{picture}(40,60)(0,0)
\put(20,0){\line(0,1){60}} \linethickness{.3pt}
\qbezier(20,20)(21.7,19.96)(22.16,21.56)
\qbezier(22.16,21.56)(22.44,23.09)(24.07,23.24)
\qbezier(24.07,23.24)(25.76,23.58)(25.77,25.19)
\qbezier(25.77,25.19)(25.55,26.69)(27,27.42)
\qbezier(27,27.42)(28.35,28.47)(27.5,29.89)
\qbezier(27.5,29.89)(26.41,31.02)(27.06,32.41)
\qbezier(27.06,32.41)(27.45,33.98)(25.86,34.67)
\qbezier(25.86,34.67)(24.26,35)(24.19,36.65)
\qbezier(24.19,36.65)(23.95,38.24)(22.25,38.36)
\qbezier(22.25,38.36)(20.6,38.37)(20,40)
\put(12,3){$p$} \put(10,52){$p'$} \put(12,27){$q$} \put(30,27){$k$}
\end{picture}
\qquad\qquad
\begin{picture}(40,60)(0,0) \linethickness{.3pt}
\multiput(20,0)(0,40){2}{\EVRwa{5}}
\qbezier(20,40)(10,30)(20,20) \qbezier(20,40)(30,30)(20,20)
\end{picture}
\qquad\qquad
\begin{picture}(60,60)(0,0) \linethickness{.3pt}
\put(20,20){\line(2,1){20}} \put(20,40){\line(2,-1){20}}
\put(20,20){\line(-1,-3){7}} \put(20,40){\line(-1,3){7}}
\put(19.5,20){\EVRwa{5}}
\put(47.5,-1){\EZwRwa{8} }
\end{picture}
\qquad\qquad
\begin{picture}(60,60)(0,0) \linethickness{.3pt}
\put(20,20){\line(-1,-3){7}} \put(20,40){\line(-1,3){7}}
\put(40,20){\line(1,-3){7}} \put(40,40){\line(1,3){7}}
\put(20,20){\line(0,1){20}} \put(40,20){\line(0,1){20}}
\multiput(20,20)(0,20){2}{\EHRwa{5}}
\end{picture}
\qquad\qquad\text{etc.}
$$
Doing the calculation in the first instance (say) one has a contribution
to the scattering matrix which, apart from constant factors,
is given by the integral
$$\int\!\!\dO^4\kk\:\frac{\d(-\pp+\qq+\kk)\,\d(-\qq-\kk+\pp')}%
{4\,\Eo_m(\qs)\,|\ks|\,(-\kk_0+|\ks|-\iO\,\e)}~.$$
Now if this were a well-defined distribution $\phi$ in two variables,
then $\bang{u\,\phi\,v}$ should be a (finite) number,
but it is immediate to check that it is not.
Similar results are found in the other cases.
\section{Electron and positron free states}
\label{S:Electron and positron free states}
The 4-spinor bundle is a complex vector bundle $\W\to\M$
with 4-dimensional fibres,
endowed with a \emph{scaled Clifford morphism} (\emph{Dirac map})
$$\g:\TO\M\to\LL\tn\End(\W):v\mapsto\g[v]$$
over $\M$ and a Hermitian metric $\kO$ on the fibres fulfilling
$$\kO(\g[v]\phi,\psi)=\kO(\phi,\g[v]\psi)~,
\quad v\in\TO_{\!x}\M~,~~\phi,\psi\in\W_{\!\!x}~,~~x\in\M.$$
Then $\kO$ (which yields the \emph{Dirac adjoint} anti-isomorphism
$\psi\mapsto\kO^\fl(\psi)$\,,
usually denoted as $\psi\mapsto\bar\psi$)
turns out to have the signature \hbox{$(+,+,-,-)$}.
If $p:\M\to\Pm$ then
$$\W=\W^+_{\!\!p}\dir{\M}\W^-_{\!\!p}~,\quad
\W^\pm_{\!\!p}:=\Ker(\g[p^\#]\mp m)~,$$
where $p^\#\equiv g^\#(p):\M\to\LL^{-2}\tn\TO\M$
is the contravariant form of $p$\,.
The restrictions of $\kO$ to these two subbundles
turn out to have the signatures
\hbox{$(+,+)$} and \hbox{$(-,-)$}\,, respectively.

Now for each $m\in\{0\}\cup\LL^{-1}$ one is led to consider
the 2-fibred bundles $\W_{\!\!m}^\pm\to\Pm\to\M$ defined by
$$\W_{\!\!m}^\pm:=\bigsqcup_{p\in\Pm}\W_{\!\!p}^\pm\subset\Pm\cart{\M}\W~.$$

The 4-spinor bundle is also endowed\footnote{See~\cite{CJ97a,C00b}
for a review of the geometry 4-spinors and 2-spinors
for electrodynamics and other field theories.}
with a \emph{spinor connection} $\Cs$,
strictly related to the spacetime connection,
such that $\g$ and $\kO$ are covariantly constant.
It is easy to see\footnote{
If $p:\M\to\Pm$ and $\psi:\M\to\W$ are parallely transported
along some curve in $\M$,
then \hbox{$\g[p^\#]\psi\mp m\,\psi$} is also parallely transported
along the same curve;
so it vanishes along the curve if it vanishes at any one point of the curve.}
that $\G_{\!\!m}$ and $\Cs$ determine projectable connections
of $\W_{\!\!m}^\pm\to\Pm\to\M$.

Now if $\T\subset\M$ is a detector,
then a free one-electron state is defined to be
a covariantly constant section $\T\to\W_{\!\!m}^+$\,.
Namely, a free one-electron state is determined
by a covariantly constant section $p:\T\to\Pm$
and by a covariantly constant section $\psi:\T\to\W$
such that $\psi(t)\in\W_{\!\!p(t)}^+$ for each $t\in\T$.
On the other hand, a free one-positron state will be represented as
a covariantly constant section $\T\to\Wc_{\!\!m}^-$\,.\footnote{
If $\V$ is a finite-dimensional complex vector space
then its \emph{conjugate space} can be defined
as $\Vc:=\V^{\lin\alin}\cong\V^{\alin\lin}$,
where $\Vl$ and $\V^{\alin}$ are, respectively,
the $\CC$-dual and antidual spaces,
that is the spaces of linear and antilinear maps $\V\to\CC$\,.
There is an anti-isomorphism $\V\to\Vc:v\mapsto\bar v$\,.
The indices relative to a conjugate basis are distinguished by a dot.}
For brevity, these 2-fibred bundles and the related 1-particle state
quantum bundles (of vector-valued generalized half-densities)
are denoted as
\begin{align*}
& \F\equiv\W_{\!\!m}^+\to\Pm\to\T~,
&& \Ft\equiv\Wc_{\!\!m}^-\to\Pm\to\T~,\\[6pt]
& \FC^1:=\DCh_{\!\Tt}(\Pm\,,\F)\equiv\DCh_{\!\Tt}(\Pm\,,\W_{\!\!m}^+)~,
&& \td\FC{}^1:=\DCh_{\!\Tt}(\Pm\,,\Ft)\equiv\DCh_{\!\Tt}(\Pm\,,\Wc_{\!\!m}^-)~.
\end{align*}

In order to introduce appropriate generalized frames
for free electron and positron states,
one needs, for each $p\in(\Pm)_\Tt$\, a frame
$$\bigl(\uu\!_\sA(p)\,,\,\vv\!_\sA(p)\bigr)~,\quad{\scriptstyle A}=1,2~$$
of $\W_{\!\!p}$ which is adapted to the splitting
$\W_{\!\!p}=\W_{\!\!p}^+\oplus\W_{\!\!p}^-$\,.
A consistent choice can be made by extending usual procedure
of the flat inertial case.\footnote{
At some point in $\T$ one fixes a spinor frame adapted to the splitting
determined by the unit vector $\t_0$\,,
and Fermi transports it along $\T$\,;
then, in each fibre, one takes the unique boost sending $\t_0$ to $p^\#/m$\,;
up to sign (which can be fixed by continuity) this boost
transforms the given spinor frame to the desired one.}
Now one gets the generalized frames
\begin{align*}
& \Asf_{p\sA}:=\Asf_p\tn\uu\!_\sA(p)
:\T\to\FC^1\equiv\DCh{}_{\!\Tt}(\Pm\,,\W_{\!\!m}^+)~,\\[6pt]
& \Csf_{p\cA}:=\Asf_p\tn\bvv\!_\cA(p)
:\T\to\td\FC{}^1\equiv\DCh{}_{\!\Tt}(\Pm\,,\Wc_{\!\!m}^-)~,
\end{align*}
respectively for electrons and positrons.

An important technical result, which is proved by elementary linear algebra,
is
\begin{align*}
& \uu\!_\sA(p)\tn\uu^\sA(p)=\oh\,(\id+\g[p^\#/m]):\W\to\W^+_{\!\!\!p}~,
\\[6pt]
& \vv\!_\sA(p)\tn\vv^\sA(p)=\oh\,(\id-\g[p^\#/m]):\W\to\W^-_{\!\!\!p}~.
\end{align*}
\section{Photon free states}\label{S:Photon free states}
For brevity, henceforth I will use the shorthand $\H\equiv\LL^{-1}\tn\TO\M$,
so that $\H^*\equiv\LL\tn\TS\M$
and the spacetime metric $g$ is an \emph{unscaled}
(\ie\ `confomally invariant') Lorentz metric in the fibres of $\H\to\M$\,.

Remember that $\Pz\subset\TS\M$
denotes the subbbundle over $\M$ of future null half-cones
in the fibres of $\TS\M$.
Consider the 2-fibred bundle $\H_{\!\sst0\bot}\to\Pz\to\M$
whose fibre over any $k\in(\Pz)_x$\,, $x\in\M$,
is the 3-dimensional real vector space
$$(\H_{\!\sst0\bot})_k:=\{\a\in\H^* : g^\#(\a,k)=0\}~.$$
Then $\H_{\!\sst0\bot}\subset\Pz\cart{\M}\H^*$
(but note that $\H_{\!\sst0\bot}$ itself
is \emph{not} a `semi-trivial' bundle
of the type $\Pz\cart{\M}\Z$\,).
Next, consider the real vector bundle $\B_{\!\RRr}^*\to\Pz$ whose fibre over
any $k\in\Pz$ is the (2-dimensional) quotient space
$$(\B_{\!\RRr})^*_k:=(\H_{\!\sst0\bot})_k / \bang{k}
\equiv \bang{k}^\sbot/\bang{k}~,$$
where $\bang{k}$ denotes the vector space generated by $k$\,.
Moreover, $\B_{\!\RRr}\equiv\B_{\!\RRr}^{**}$ can be equivalently introduced
by a similar contravariant construction.

It turns out that the spacetime metric `passes to the quotient',
so it naturally determines
a negative metric $g_\Bb$ in the fibres of $\B_{\!\RRr}\to\Pz$\,,
as well as a `Hodge' isomorphism ${*}_{\!\Bb}$
which can be characterized through the rule\footnote{
$k\we\b$ is well defined because $\b$ is an equivalence class of covectors
differing for a term proportional to $k$\,.}
$${*}(k\we\b)=-k\we({*}_{\!\Bb}\b)~.$$

Now define the \emph{optical bundle}\footnote{
In the literature this term is often used in a somewhat different
(but related) sense,
denoting a vector bundle over $\M$ associated with the choice
of a congruence of null lines~\cite{Nu}.}
to be the 2-fibred bundle
$$\B:=\CC\tn\B_{\!\RRr}\to\Pz\to\M~.$$
This has the canonical splitting
$$\B=\B^+ \dir{\Pz} \B^-~,$$
where the fibres of the subbundles $\B^\pm\to\Pz$ are defined to be
the eigenspaces of $-\iO\,{*}_{\!\Bb}$ with eigenvalues $\pm1$
(and turn out to be 1-dimensional complex $g_\Bb$-null subspaces).

Let $u:\M\to\TO\M$ be any given `observer',
\ie\ a unit timelike vector field on $\M$\,;
through $u$ one can identify $\B_\RRr\to\Pz$ with
$\H_{\!\sst0\bot}\cap\bang{u}^\sbot\to\Pz$ (`radiation gauge').
Take any $k\in(\Pz)_x$\,, $x\in\M$,
and let $(\ee_\l)$\,, $\l=0,1,2,3$,
be an orthonormal basis of $\H_{\!x}$ such that $\ee_0\equiv u(x)$
and $k^\#\propto\ee_0+\ee_3$\,;
then the basis
$$(\bb_1\,,\bb_2)\equiv(\bb_+\,,\bb_-)
:=\Bigl(\osq\,(\ee_1+\iO\,\ee_2)~,~\osq\,(\ee_1-\iO\,\ee_2)\Bigr)
\subset\CC\tn(\H_{\!\sst0\bot}\cap\bang{u}^\sbot)$$
is \emph{adapted} to the splitting of $\B_k$\,,
\ie\ $\bb_\pm\in\B_k^\pm$\,.
In this way, locally one can construct smooth frames of
$\CC\tn\H\cart{\M}\Pz$\,,
and smooth frames of $\B\to\Pz$ which are adapted to its splitting.

Let now $\{\Bsf_k\}$ be the generalized frame of the quantum bundle
$\PC_{\!\!\sst0}\to\M$ defined as usual;
then one gets a generalized frame
$$\{\Bsf_{\k\sQ}\}:=\{\Bsf_k\tn\bb_\sQ(k)\}~,\quad {\scriptstyle Q}=1,2\,,$$
of the quantum bundle
$$\BC^1:=\DCh_{\!\Mm}(\Pz\,,\B)\to\M~.$$
In particular, all the above bundles and constructions can be restricted
to a detector $\T\subset\M$.
In that case, $\ee_0$ will be chosen to be the unit future-pointing vector
tangent to $\T$.

Free asymptotic 1-photon states will be described
as covariantly constant sections $\T\to\BC^1$
(they only possess \emph{transversal polarization modes}).
Virtual photons, on the other hand, span a larger bundle;
they are described as covariantly constant sections
$$\T\to\td\BC^1\equiv\DCh{}_{\!\Tt}(\Pz\,,\CC\tn\H)~,$$
where one uses the generalized frame
$$\{\Bsf_{\k\l}\}:=\{\Bsf_k\tn\ee_\l\}~,\quad \l=0,1,2,3\,.$$
\section{Electromagnetic interaction}\label{S:Electromagnetic interaction}
The `classical electromagnetic interaction' is the 3-linear morphism
\begin{align*}
\ell\interaction&:\Wc\cart{\M}\H\cart{\M}\W\to\CC \\[6pt]
&:(\phi,b,\psi)\mapsto -e\,\kO(\phi,\g[b]\psi)
\equiv -e\,\bang{\kO^\fl\phi,\g[b]\psi}~,
\end{align*}
where $e\in\RR^+$ is the positron's charge
(a pure number in natural units).

As sketched in~\S\ref{S:Quantum interaction},
this geometric structure of the underlying classical bundles,
together with the generalized half-density $\ul\Lambda$\,,
determines the quantum interaction $-\iO\,\Hfr$\,.
A short discussion is needed in order to see how the index types
of the various terms in $\Hfr$ arise.

First, $\ell\interaction$ is extended to $\Wc\cart{\M}\H_{\!\CCc}\cart{\M}\W$,
where $\H_{\!\CCc}\equiv\CC\tn\H$.
In the fibres of the complex vector bundle $\H_{\!\CCc}\to\M$
indices are raised and lowered
through the obvious extension of the spacetime metric $g$\,,
while in the fibres of $\B\to\Pz$ one uses $g_\Bb$\,;
when an observer is chosen and one works in the radiation gauge,
the latter operation can be viewed essentially as a restriction of the former.

Now $\ell\interaction$ can be seen as a $\CC$-linear function
on the fibres of
$$\Wc\ten{\Ptru}\H_{\!\CCc}\ten{\Ptru}\W\to
\Ptru\equiv\Pm\cart{\M}\Pz\cart{\M}\Pm~.$$
Note that the isomorphism $\kO^\fl:\Wc\to\Wl$
induced by the Hermitian metric $\kO$ preserves
the splitting $\W\cart{\M}\Pm=\W_{\!\!m}^+\dir{\Pm}\W_{\!\!m}^-$\,,
namely $\kO^\fl:\Wc_{\!\!m}^\pm\to(\W_{\!\!m}^\pm)^\lin$\,.
Then
\begin{align*}
& \Pm\cart{\M}\Wc=\Wc_{\!\!m}^+\dir{\Pm}\Wc_{\!\!m}^-
\cong(\W_{\!\!m}^+)^\lin\dir{\Pm}\Wc_{\!\!m}^-~,\\[6pt]
& \Pm\cart{\M}\W=\W_{\!\!m}^+\dir{\Pm}\W_{\!\!m}^-
\cong\W_{\!\!m}^+\dir{\Pm}(\Wc_{\!\!m}^-)^\lin~.
\end{align*}
When $\Pm\cart{\M}\W$ and $\Pm\cart{\M}\Wc$ are written in this way,
the coordinate expression of
$$\ell\interaction:
\Bigl( (\W_{\!\!m}^+)^\lin\dir{\Pm}\Wc_{\!\!m}^-\Bigr)
\ten{\Ptru}\H_{\!\CCc}\ten{\Ptru}
\Bigl(\W_{\!\!m}^+\dir{\Pm}(\Wc_{\!\!m}^-)^\lin\Bigr) \to \CC $$
contains four terms with different index types;
all dotted (\ie\ `conjugated') indices, either high or low,
refer to the positron bundle $\Wc_{\!\!m}^-$ or to its dual,
while undotted indices refer to the electron bundle or to its dual.
Finally, a further extension through $g$ gives
$$\ell\interaction:
\Bigl( (\W_{\!\!m}^+)^\lin\dir{\Pm}\Wc_{\!\!m}^- \Bigr)
\ten{\Ptru}\Bigl(\H_{\!\CCc}\dir{\Pz}\H_{\!\CCc}^\lin\Bigr)\ten{\Ptru}
\Bigl(\W_{\!\!m}^+\dir{\Pm}(\Wc_{\!\!m}^-)^\lin \Bigr) \to \CC~,$$
which is the sum of \emph{eight} terms of different index types.
Explicitely, if $\b\in\H_{\!\CCc}$ then one replaces $\g[b]$
in the expression of $\ell\interaction$ with
$$\g^\#[\b]:=\g[\b\#]\equiv\g[g^\#\b]~.$$

Further objects can be obtained
by exchanging tensor factors in $\ell\interaction$\,.
However, objects only distinguished for a different order of
indices referring to different particle types are regarded as equivalent,
while in the different ordering of indices referring to the same
particle type only those terms are retained
which have the covariant indices
\emph{on the right} of the contravariant ones.

Let now $\T\subset\M$ be a detector,
consider the restrictions to $\T$ of the various quantum bundles
and the Fock bundle $\td\FC\ten{\T}\td\BC\ten{\T}\FC$\,.
At this point, one has all the ingredients needed to write down
the interaction morphism over $\T$
$$\Hfr:\td\FCo\ten{\T}\td\BCo\ten{\T}\FCo
\to\LL^{-1}\tn\td\FC\ten{\T}\td\BC\ten{\T}\FC~,$$
where the subscript circles indicate the subbundles of test elements.
The constant $\l\in\LL^{-1}$
introduced at the end of~\S\ref{S:Quantum interaction} is, here,
the electron's mass $m$\,.
Thus one finds
\begin{align*}
\frac1m\,\Hfr
&=\eO^{\iO\,(-\pp_0-\kk_0-\qq_0)\,\tt}\,\La_{\pp\cA\,\kk\l\,\qq\sB}\,
      \Csf^{\pp\cA}\tn\Bsf^{\kk\l}\tn\Asf^{\qq\sB}~+ && \Lcqs
\\[6pt]
&\quad+\eO^{\iO\,(\pp_0-\kk_0-\qq_0)\,\tt}\,
       \La\Ii{\pp\sA}{\kk\l\,\qq\sB}\,
       \Asf_{\pp\sA}\tn\Bsf^{\kk\l}\tn\Asf^{\qq\sB}~+ && \LSqs
\displaybreak[2]\\[6pt]
&\quad+\eO^{\iO\,(-\pp_0+\kk_0-\qq_0)\,\tt}\,
       \La\iIi{\pp\cA}{\kk\l}{\qq\sB}\,
       \Csf^{\pp\cA}\tn\Bsf_{\kk\l}\tn\Asf^{\qq\sB}~+ && \LcQs
\\[6pt]
&\quad+\eO^{\iO\,(\pp_0-\kk_0-\qq_0)\,\tt}\,
       \La\Ii{\pp\cA}{\kk\l\,\qq\cB}\,
       \Csf_{\pp\cA}\tn\Bsf^{\kk\l}\tn\Csf^{\qq\cB}~+ && \LCqc
\displaybreak[2]\\[6pt]
&\quad+\eO^{\iO\,(\pp_0+\kk_0-\qq_0)\,\tt}\,
       \La\Ii{\pp\sA\,\kk\l}{\qq\sB}\,
       \Asf_{\pp\sA}\tn\Bsf_{\kk\l}\tn\Asf^{\qq\sB}~+ && \LSQs
\\[6pt]
&\quad+\eO^{\iO\,(\pp_0-\kk_0+\qq_0)\,\tt}\,
       \La\IiI{\pp\sA}{\kk\l}{\qq\cB}\,
       \Asf_{\pp\sA}\tn\Bsf^{\kk\l}\tn\Csf_{\qq\cB}~+ && \LSqC
\displaybreak[2]\\[6pt]
&\quad+\eO^{\iO\,(\pp_0+\kk_0-\qq_0)\,\tt}\,
       \La\Ii{\pp\cA\,\kk\l}{\qq\cB}\,
       \Csf_{\pp\cA}\tn\Bsf_{\kk\l}\tn\Csf^{\qq\cB}~+ && \LCQc
\\[6pt]
&\quad+\eO^{\iO\,(\pp_0+\kk_0+\qq_0)\,\tt}\,
       \La^{\pp\sA\,\kk\l\,\qq\cB}\,
       \Asf_{\pp\sA}\tn\Bsf_{\kk\l}\tn\Csf_{\qq\cB}~. && \LSQC
\end{align*}
where
$$\La_{\pp\cA\,\kk\l\,\qq\sB}
=\ell_{\pp\cA\,\kk\l\,\qq\sB}\,\ul\La_{\pp\,\kk\qq}
=\ell_{\pp\cA\,\kk\l\,\qq\sB}\,
\frac{\d(\ps+\ks+\qs)}{\sqrt{8\,l^9\,\pp_0\,\kk_0\,\qq_0}}$$
and the like.
Explicitely, the $\ell$-factors are given by
\begin{align*}
&\ell_{\pp\cA\,\kk\l\,\qq\sB}=-e\,\bvv_{\pp\cA}\,\g_{\kk\l}\,\uu_{\qq\sB}~,
&& \ell\Ii{\pp\sA}{\kk\l\,\qq\sB}=-e\,\uu^{\pp\sA}\,\g_{\kk\l}\,\uu_{\qq\sB}~,
\\[6pt]
&\ell\iIi{\pp\cA}{\kk\l}{\qq\sB}=-e\,\bvv_{\pp\cA}\,\g^{\kk\l}\,\uu_{\qq\sB}~,
&&\ell\Ii{\pp\cA}{\kk\l\,\qq\cB}=-e\,\bvv_{\qq\cB}\,\g_{\kk\l}\,\bvv^{\pp\cA}~,
\\[6pt]
&\ell\Ii{\pp\sA\,\kk\l}{\qq\sB}=-e\,\uu^{\pp\sA}\,\g^{\kk\l}\,\uu_{\qq\sB}~,
&&\ell\IiI{\pp\sA}{\kk\l}{\qq\cB}=-e\,\uu^{\pp\sA}\,\g_{\kk\l}\,\bvv^{\qq\cB}~,
\\[6pt]
&\ell\Ii{\pp\cA\,\kk\l}{\qq\cB}=-e\,\bvv_{\qq\cB}\,\g^{\kk\l}\,\bvv^{\pp\cA}~,
&&\ell^{\pp\sA\,\kk\l\,\qq\cB}=-e\,\uu^{\pp\sA}\,\g^{\kk\l}\,\bvv^{\qq\cB}~,
\end{align*}
where $\ee_{\kk\l}\equiv\ee_\l(\kk)$\,,
$\ee^{\kk\l}\equiv\ee^\l(\kk)$ denotes its dual frame of $\H_{\!\CCc}^\lin$\,.
Morever, $\g_{\kk\l}\equiv\g[\ee_{\kk\l}]$ and
$\g^{\kk\l}\equiv\g^\#[\ee^{\kk\l}]$\,.

In the above elementary diagrams time runs upwards;
so, lines entering the vertex from below represent absorbed particles,
lines entering from above represent created particles;
electron lines are labelled by up arrows,
positron lines are labelled by down arrows,
and photon lines are wavy.
\section{QED}\label{S:QED}
In this section I will show how two-point interactions\footnote{
One-point interactions in QED are nearly obvious at this stage.}
give rise to scattering matrix contributions which, at least formally,
have the same expressions as in standard treatments;
these expressions are the so-called \emph{propagators}
of the particles in momentum space.
In the flat inertial case one recovers standard results.

Consider a second order process in which the initial and final states
both contain one electron and one photon.
One has two types of diagrams,
and for each type on turn two subtypes can be distinguished,
according to the time order of the vertices:
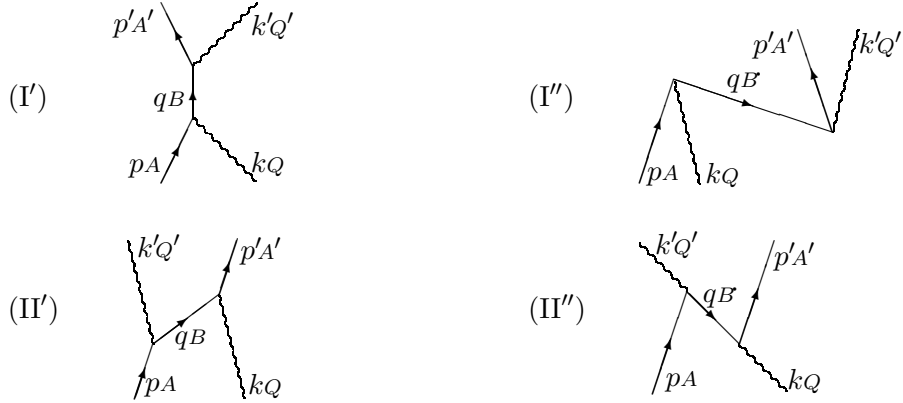
\begin{align*}
&\begin{picture}(60,70)(0,0)
\put(-40,30){(${\mathrm{I}}'$)}
\put(30,45){\line(-1,2){12}} \put(30,45){\vector(-1,2){7}}
\put(18,1){\line(1,2){12}}  \put(18,1){\vector(1,2){7}}
\put(30,25){\line(0,1){20}} \put(30,25){\vector(0,1){11}}
\linethickness{.3pt}
\put(30,45){\EDeLwa{8}} \put(53.8,1.5){\EDwLwa{8}}
\put(6,5){$p{\scriptstyle A}$} \put(0,58){$p'\!{\scriptstyle A}'$}
\put(52,5){$k{\scriptstyle Q}$} \put(52,57){$k'\!{\scriptstyle Q}'$}
\put(15,30){$q{\scriptstyle B}$}
\end{picture}
&\hskip3cm&\begin{picture}(100,60)(0,0)
\put(-40,30){(${\mathrm{I}}''$)}
\put(15,40){\line(-1,-3){13}} \put(2,1){\vector(1,3){8}}
\put(75,20){\line(-1,3){13}} \put(75,20){\vector(-1,3){8}}
\put(15,40){\line(3,-1){60}} \put(15,40){\vector(3,-1){30}}
\linethickness{.3pt}
\put(24.8,.4){\EZwRwa{10}} \put(75,20){\EZeRwa{10}}
\put(5,2){$p{\scriptstyle A}$} \put(45,50){$p'\!{\scriptstyle A}'$}
\put(27,1){$k{\scriptstyle Q}$} \put(85,50){$k'\!{\scriptstyle Q}'$}
\put(35,37){$q{\scriptstyle B}\.$}
\end{picture}
\displaybreak[2]\\[12pt]
&\begin{picture}(60,60)(0,0)
\put(-40,30){(${\mathrm{II}}'$)}
\put(15,20){\line(-1,-3){7}} \put(8,-1){\vector(1,3){4}}
\put(15,20){\line(4,3){25}} \put(15,20){\vector(4,3){13}}
\put(40,39){\line(1,3){7}} \put(40,39){\vector(1,3){4}}
\linethickness{.3pt}
\put(15,20){\EZwRwa{10}} \put(49.5,-.5){\EZwRwa{10}}
\put(12,2){$p{\scriptstyle A}$} \put(48,52){$p'\!{\scriptstyle A}'$}
\put(9,53){$k'\!{\scriptstyle Q}'$} \put(51,1){$k{\scriptstyle Q}$}
\put(23,20){$q{\scriptstyle B}$}
\end{picture}
&&
\begin{picture}(70,60)(0,0)
\put(-40,30){(${\mathrm{II}}''$)}
\put(20,40){\line(1,-1){20}} \put(20,40){\vector(1,-1){11}}
\put(20,40){\line(-1,-3){13}} \put(7,1){\vector(1,3){7}}
\put(40,20){\line(1,3){13}} \put(40,20){\vector(1,3){7}}
\linethickness{.3pt}
\put(20,40){\EDwRwa{6}}
\put(57.5,2){\EDwRwa{6}}
\put(12,4){$p{\scriptstyle A}$} \put(53,50){$p'\!{\scriptstyle A}'$}
\put(8,54){$k'\!{\scriptstyle Q}'$} \put(58,3){$k{\scriptstyle Q}$}
\put(26,36){$q{\scriptstyle B}\.$}
\end{picture}
\end{align*}
Here, external photon lines are labelled by an index ${\scriptstyle Q}=1,2$
referring to the classical frame $(\bb_{\kk\sQ})$ of the bundle $\B\to\Pz$
of transversal polarization modes;
4-momenta are indicated by letters $p$, $q$, $k$ etcetera.

First, consider the diagram labelled as (I').
Following the usual procedure one finds
\begin{align*}
&(\Scal_{\mathrm I'})\Ii{\pp'\!\sA' \kk'\!\sQ'}{\pp\sA\,\kk\sQ}
=-m^2\int\limits_{-\infty}^{+\infty}\dt_2\int\limits_{-\infty}^{+\infty}\dt_1
\int\dO^3\qs\:\he(\tt_2-\tt_1)\,
\Bigl(\sum_{\sB=1}^2\ell\IiI{\pp'\!\sA'}{\qq\sB}{\kk'\!\sQ'}\,
\ell\iIi{\pp\sA}{\qq\sB}{\kk\sQ}\Bigr)~\cdot
\\[8pt] &\hskip2.5cm
\cdot~\frac{\d(\ps'-\qs+\ks')\,\d(-\ps+\qs-\ks)}%
{l^6\,\sqrt{16\,\pp_0'\,\kk_0'\,\pp_0\,\kk_0}\:2\,\qq_0}\cdot
\eO^{\iO(-\pp_0+\qq_0-\kk_0)\tt_1}\,\eO^{\iO(\pp_0'-\qq_0+\kk_0')\tt_2}~.
\end{align*}
Note that the summation over ${\scriptstyle B}$\,, here,
is to be performed before all other operations:
it must be performed before integration over $\qq$ because
the index ${\scriptstyle B}$ `resides' over $\qq$\,,
and before the transformation of the integral into a 4-dimensional
one because the index ${\scriptstyle B}$ cannot reside
over an off-shell momentum.
Hence one considers, for \emph{fixed} $\qq$\,,
\begin{align*}
&\sum_{\sB=1}^2\ell\IiI{\pp'\!\sA'}{\qq\sB}{\kk'\!\sQ'}\,
\ell\iIi{\pp\sA}{\qq\sB}{\kk\sQ}=
e^2\,\sum_{\sB=1}^2(\uu^{\pp'\!\sA'}\,\g^{\kk'\!\sQ'}\,\uu_{\qq\sB})\:
(\uu^{\qq\sB}\,\g_{\kk\sQ}\,\uu_{\pp\sA})=\\[6pt]
&\quad
=e^2\,\sum_{\sB=1}^2\uu^{\pp'\!\sA'}\comp\g^{\kk'\!\sQ'}\comp
(\uu_{\qq\sB}\tn\uu^{\qq\sB})\comp\g_{\kk\sQ}\,\uu_{\pp\sA}=\\[6pt]
&\quad
=e^2\,\uu^{\pp'\!\sA'}\comp\g^{\kk'\!\sQ'}\comp(\id+\tfrac1m\,\g^\#[\qq])
\comp\g[\bb_{\kk\sQ}]\,\uu_{\pp\sA}~,
\end{align*}
since $\sum_{\sB=1}^2\uu_{\qq\sB}\tn\uu^{\qq\sB}$\,, for fixed $\qq$\,,
is just the projection $\id{+}\tfrac1m\,\g^\#[\qq]:\W\to\W^+_{\!\!\qq}$\,.
Furthermore, observe that for $\qq\in\Pm$ one has
$$\id+\tfrac1m\,\g^\#[\qq]
=\tfrac1m\,(m+\Eo_m(\qs)\,\g^0+\g^\#[\qs])~,\quad
\Eo_m(q_\sbo):=\sqrt{m^2+|\qs|^2}~.$$
Now when one performs the usual trick for transforming the integral
into a 4-dimensional one, the above factor remains unchanged, so that
\begin{align*}
&(\Scal_{\mathrm I'})\Ii{\pp'\!\sA' \kk'\!\sQ'}{\pp\sA\,\kk\sQ}=
\frac{2\pi\iO\,m\,e^2}{l^6\,\sqrt{16\,\pp_0'\,\kk_0'\,\pp_0\,\kk_0}}\,
\uu^{\pp'\!\sA'}\comp\g^{\kk'\!\sQ'}\comp \\[8pt]
&\qquad\comp\Bigl(
\int\dO^4\qq\,
\frac{\d(-\pp-\kk+\qq)\,\d(-\qq+\pp'+\kk')}%
{2\,\Eo_m(\qs)\,(-\qq_0+\Eo_m(\qs)-\iO\,\e)}\,
\bigl(m+\Eo_m(\qs)\,\g^0+\g^\#[\qs]\bigr)\Bigr)
\comp\g_{\kk\sQ}\,\uu_{\pp\sA}~.
\end{align*}

Next, consider the diagram labelled as (I'').
Like in the scalar case,
the different time order of the vertices yields different signs
in the arguments of the Dirac deltas.
The classical Lagrangian yields now a further difference, since
$$ \sum_{\cB=1}^2\ell^{\pp'\!\sA'\,\kk'\!\sQ'\,\qq\cB}\,
\ell_{\pp\sA\,\kk\sQ\,\qq\cB}
=e^2\,\uu^{\pp'\!\sA'}\comp\g^{\kk'\!\sQ'}\comp(\id-\tfrac1m\,\g^\#[\qq])
\comp\g_{\kk\sQ}\,\uu_{\pp\sA}~.$$
Then one finds
\begin{align*}
&(\Scal_{\mathrm I''})\Ii{\pp'\!\sA' \kk'\!\sQ'}{\pp\sA\,\kk\sQ}=
\frac{2\pi\iO\,m\,e^2}{l^6\,\sqrt{16\,\pp_0'\,\kk_0'\,\pp_0\,\kk_0}}\,
\uu^{\pp'\!\sA'}\comp\g^{\kk'\!\sQ'}\comp \\[8pt]
&\qquad\comp\Bigl(
\int\dO^4\qq\,
\frac{\d(-\ps-\ks-\qs)\,\d(\qs+\ps'+\ks')}%
{2\,\Eo_m(\qs)\,(\qq_0+\Eo_m(\qs)-\iO\,\e)}
\d(-\pp_0-\kk_0+\qq_0)\,\d(-\qq_0+\pp_0'+\kk_0')\cdot
\\[8pt]
& \hspace{7cm}
\cdot\bigl(m-\Eo_m(\qs)\,\g^0-\g^\#[\qs]\bigr)\Bigr)
\comp\g_{\kk\sQ}\,\uu_{\pp\sA}~.
\end{align*}
In order to simplify
$(\Scal_{\mathrm I})\Ii{\pp'\!\sA' \kk'\!\sQ'}{\pp\sA\,\kk\sQ}=
(\Scal_{\mathrm I'}+\Scal_{\mathrm I''})
\Ii{\pp'\!\sA' \kk'\!\sQ'}{\pp\sA\,\kk\sQ}$
one has to make the integration variable change $\qs\to{-\qs}$
in the second contribution, so that the $\d$-factors are the same.
Eventually,
\begin{align*}
&(\Scal_{\mathrm I})\Ii{\pp'\!\sA' \kk'\!\sQ'}{\pp\sA\,\kk\sQ}=
\frac{-2\pi\iO\,m\,e^2}{l^6\,\sqrt{16\,\pp_0'\,\kk_0'\,\pp_0\,\kk_0}}\,
\uu^{\pp'\!\sA'}\comp\g^{\kk'\!\sQ'}\comp \\[8pt]
&\qquad\comp\Bigl(
\int\dO^4\qq\,\d(-\pp-\kk+\qq)\,\d(-\qq+\pp'+\kk')\cdot
\frac{m+\g^\#[\qq]}{g(\qq,\qq)-m^2+\iO\,\e}\Bigr)\comp
\g_{\kk\sQ}\,\uu_{\pp\sA}~,
\end{align*}
which contains the \emph{electron propagator}, namely the distribution
$$\lim_{\e\to0^+}\frac{-m-\g^\#[\qq]}{g(\qq,\qq)-m^2+\iO\,\e}~.$$
The \emph{positron propagator}
$$\lim_{\e\to0^+}\frac{-m+\g^\#[\qq]}{g(\qq,\qq)-m^2+\iO\,\e}~.$$
is found by a similar procedure.

\medbreak

Next, consider the diagrams
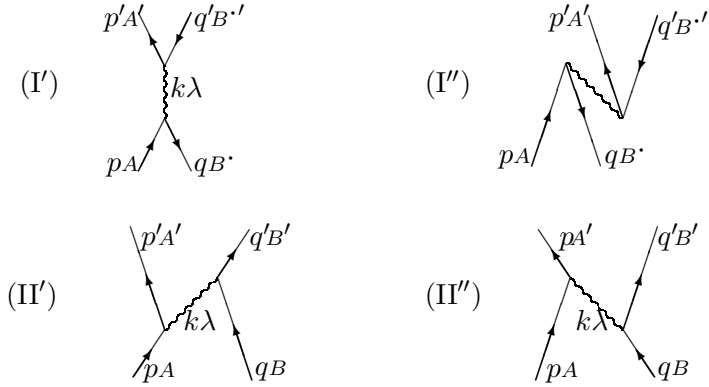
\begin{align*}
& \begin{picture}(40,60)(0,0)
\put(-35,30){(${\mathrm I}'$)}
\put(10,60){\line(1,-2){10}} \put(20,40){\vector(-1,2){6}}
\put(20,40){\line(1,2){10}} \put(30,60){\vector(-1,-2){6}}
\put(10,0){\line(1,2){10}} \put(10,0){\vector(1,2){6}}
\put(20,20){\line(1,-2){10}} \put(20,20){\vector(1,-2){6}}
\linethickness{.3pt} \put(20,20){\EVRwa{5}}
\put(-2,0){$p{\scriptstyle A}$} \put(-3,55){$p'\!{\scriptstyle A}'$}
\put(31,0){$q{\scriptstyle B}\,\.$} \put(31,55){$q'\!{\scriptstyle B}\,\.{\,}'$}
\put(22,28){$k\l$}
\end{picture}
&&
\begin{picture}(60,60)(0,0)
\put(-35,30){(${\mathrm I}''$)}
\put(18.5,41){\line(-1,-3){13}} \put(5.5,2){\vector(1,3){7}}
\put(18.5,41){\line(1,-3){13}} \put(18.5,41){\vector(1,-3){7}}
\put(40,20){\line(-1,3){13}} \put(40,20){\vector(-1,3){7}}
\put(40,20){\line(1,3){13}} \put(53,59){\vector(-1,-3){7}}
\linethickness{.3pt} \put(39.5,20){\EDwLwa{7}}
\put(-7,3){$p{\scriptstyle A}$} \put(12,55){$p'\!{\scriptstyle A}'$}
\put(33,3){$q{\scriptstyle B}\,\.$}
\put(53,52){$q'\!{\scriptstyle B}\,\.{\,}'$}
\end{picture}
\displaybreak[2]\\[12pt]
&\begin{picture}(60,60)(0,0)
\put(-40,30){(${\mathrm{II}}'$)}
\put(20,20){\line(-1,3){13}} \put(20,20){\vector(-1,3){7}}
\put(20,20){\line(-2,-3){12}} \put(8,2){\vector(2,3){7}}
\put(40,40){\line(2,3){12}} \put(40,40){\vector(2,3){7}}
\put(40,40){\line(1,-3){13}} \put(53,1){\vector(-1,3){7}}
\linethickness{.3pt} \put(20,20){\ODeRwa{6}}
\put(12,2){$p{\scriptstyle A}$} \put(11,52){$p'\!{\scriptstyle A}'$}
\put(54,3){$q{\scriptstyle B}$} \put(52,52){$q'\!{\scriptstyle B}'$}
\put(27,20){$k\l$}
\end{picture}
&&
\begin{picture}(60,60)(0,0)
\put(-35,30){(${\mathrm{II}}''$)}
\put(20,40){\line(-2,3){12}} \put(20,40){\vector(-2,3){7}}
\put(20,40){\line(-1,-3){13}} \put(7,1){\vector(1,3){7}}
\put(40,20){\line(2,-3){12}} \put(52,2){\vector(-2,3){7}}
\put(40,20){\line(1,3){13}} \put(40,20){\vector(1,3){7}}
\linethickness{.3pt} \put(39.5,20){\ODwRwa{6}}
\put(11,2){$p{\scriptstyle A}$} \put(16,52){$p\!{\scriptstyle A}'$}
\put(53,3){$q{\scriptstyle B}$} \put(53,52){$q'\!{\scriptstyle B}'$}
\put(22,20){$k\l$}
\end{picture}
\end{align*}
From case (II) one can obtain two further similar cases
by inverting one or both fermion paths.
In all cases the calculation is essentially the same;
case (II) is somewhat simpler notationally since it has no dotted indices.
Diagram (II') yields the summation
$$\sum_{\l=0}^3 \ell\Ii{\pp'\!\sA'\kk\l}{\pp\sA}\,
\ell\Ii{\qq'\!\sB'}{\kk\l\,\qq\sB}=
e^2\,\sum_{\l=0}^3
(\uu^{\pp'\!\sA'}\,\g^{\kk\l}\,\uu_{\pp\sA})\:
(\uu^{\qq'\!\sB'}\,\g_{\kk\l}\,\uu_{\qq\sB})$$
over the internal polarization degrees of freedon of the photon;
the generalized index $\kk$ is kept fixed (no summation on it).
Here $\g_{\kk\l}\equiv\g[\ee_{\kk\l}]$
and $\g^{\kk\l}\equiv\g^\#[\ee^{\kk\l}]$\,.

In order to handle the above expression conveniently,
look at the Dirac map $\g$ as a linear morphism $\H\to\W\tn\Wl$\,,
so that
$$\g[y]\tn\g[y]\in(\W\tn\Wl)\tn(\W\tn\Wl)=\LO(\Wl\tn\W,\W\tn\Wl)~,
\quad y\in\H~.$$
One then finds
$$\g^{\kk\l}\tn\g_{\kk\l}=g^{\l\m}\,\g_{\kk\l}\tn\g_{\kk\m}
=g_{\l\m}\,\g^{\kk\l}\tn\g^{\kk\m}:\Wl\tn\W\to\W\tn\Wl~.$$
Moreover, mote that the generalized index $\kk$ in the above expression
\emph{can be dropped},
since the described object is independent of the frame in which it is written,
namely
$$\g^{\kk\l}\tn\g_{\kk\l}=\g^{\kk'\l'}\tn\g_{\kk'\l'}
\equiv\g^\l\tn\g_\l~.$$

Now the previously considered summation over the virtual photon's
polarization states can be rewritten as
$$\sum_{\l=0}^3 \ell\Ii{\pp'\!\sA'\kk\l}{\pp\sA}\,
\ell\Ii{\qq'\!\sB'}{\kk\l\,\qq\sB}
=e^2\,g_{\l\m}\,(\uu^{\pp'\!\sA'}\tn\uu_{\pp\sA})\comp(\g^\l\tn\g^\m)\comp
(\uu^{\qq'\!\sB'}\tn\uu_{\qq\sB})~.$$
From diagram (II') one gets, similarly,
$$\sum_{\l=0}^3 \ell\Ii{\pp'\!\sA'}{\kk\l\,\pp\sA}\,
\ell\Ii{\qq'\!\sB' \kk\l}{\qq\sB}
=e^2\,g_{\l\m}\,(\uu^{\pp'\!\sA'}\tn\uu_{\pp\sA})\comp(\g^\l\tn\g^\m)\comp
(\uu^{\qq'\!\sB'}\tn\uu_{\qq\sB})~.$$
Thus, eventually,
the contraction over the internal states of the virtual photon
gives the same result in the two subcases (II') and (II'')
differing for the time ordering of the interaction.
The fact that the generalized index $\kk$ disappears in this operation
implies that the photon propagator is simply the scalar massless one
tensorialized by the spacetime metric, that is
\begin{align*}
&(\Scal_{\mathrm{II}})\Ii{\pp'\!\sA' \qq'\!\sB'}{\pp\sA\,\qq\sB}=
\frac{-2\pi\iO\,m^2\,e^2}{l^6\,\sqrt{16\,\pp_0'\,\qq_0'\,\pp_0\,\qq_0}}\,
(\uu^{\pp'\!\sA'}\tn\uu_{\pp\sA})\comp\g^\l\comp \\[8pt]
&\qquad\comp\Bigl(
\int\dO^4\kk\,
\d(-\pp-\kk+\qq)\,\d(-\qq+\pp'+\kk')\cdot
\frac{g_{\l\m}}{g(\kk,\kk)+\iO\,\e}
\Bigr)\cdot
\g^\m(\uu^{\qq'\!\sB'}\tn\uu_{\qq\sB})~.
\end{align*}


\begin{thebibliography}{HCMN95}
%
\bibitem[BLT75]{BLT}
Bogolubov, N.N., Logunov, A.A.\ and Todorov, I.T.:
\emph{Introduction to Axiomatic Quantum Field Theory},
Benjamin, Reading (1975).
%
\bibitem[CK95]{CK95}
Cabras, A.\ and Kol\'a\v{r}, I.:
`Connections on some functional bundles',
Czech.\ Math.\ J.\ {\bf 45}, 120 (1995), 529--548.
%
\bibitem[CJ97a]{CJ97a}
Canarutto, D.\ and Jadczyk A.:
`Fundamental geometric structures
for the Dirac equation in General Relativity',
Acta Appl.\ Math.\ {\bf 50} N.1 (1998), 59--92.
%
\bibitem[C00a]{C00a}
Canarutto, D.:
`Smooth bundles of generalized half-densities',
Archivum Mathematicum, Brno, {\bf 36} (2000), 111--124.
%
\bibitem[C00b]{C00b}
Canarutto, D.:
`Two-spinors, field theories and geometric optics in curved spacetime',
Acta Appl.\ Math.\ {\bf 62} N.2 (2000), 187--224.
%
\bibitem[C04a]{C04a}
Canarutto, D.:
`Connections on distributional bundles',
Rend.\ Semin.\ Mat.\ Univ.\ Padova {\bf 111} (2004), 71--97.
%
\bibitem[C04b]{C04b}
Canarutto, D.:
`Quantum connections and quantum fields',
Rend.\ Ist.\ Mat.\ Univ.\ Trieste, {\bf 36} (2004), 1--21.
%
\bibitem[CJM95]{CJM}
Canarutto, D., Jadczyk A.\ and Modugno, M.:
`Quantum mechanics of a spin particle
in a curved spacetime with absolute time',
Rep.\ Math.\ Phys.\ {\bf 36} (1995), 95--140.
%
\bibitem[JM02]{JM}
Jany{\v s}ka, J.\ and Modugno, M.:
`Covariant Schr\"odinger operator',
J.\ Phys.\ A {\bf 35} (2002), 8407--8434.
%
\bibitem[Fr82]{Fr}
Fr\"olicher, A.:
\emph{Smooth structures},
Lecture Notes in Mathematics {\bf 962}, Springer-Verlag (1982) 69--81.
%
\bibitem[FK88]{FK}
Fr\"olicher, A.\ and Kriegl, A.:
\emph{Linear spaces and differentiation theory},
John Wiley \& sons (1988).
%
\bibitem[KM97]{KM}
Kriegl, A.\ and Michor, P.:
\emph{The convenient setting of global analysis},
American Mathematical Society (1997).
%
\bibitem[MK98]{MK}
Modugno, M.\ and Kol\'a\v{r}, I.:
\emph{The Fr\"olicher-Nijenhuis bracket on some functional spaces},
Ann.\ Pol.\ Math.\ {\bf 68} (1998), 97--106.
%
\bibitem[Nu96]{Nu}
Nurowski, P.:
Optical geometries and related structures,
J.\ Geom.\ Phys.\ {\bf 18} (1996), 335--348.
%
\bibitem[Sc66]{Sc}
Schwartz, L.:
\emph{Th\'eorie des distributions},
Hermann, Paris (1966).
%
\end{thebibliography}
\end{document}